\documentclass[fleqn,usenatbib]{mnras}

\usepackage{newtxtext,newtxmath}

\usepackage[T1]{fontenc}
\usepackage{ae,aecompl}
\usepackage{soul}

\usepackage{graphicx}	

\newcommand{\heii}{He{\sc ii}}
\newcommand{\civ}{C{\sc iv}}
\newcommand{\Lya}{Ly$\alpha$}

\newcommand{\mpyr}{$M_\odot\,\rm yr^{-1}$}
\newcommand{\kmps}{$\rm km\,s^{-1}$}
\newcommand{\fcgs}{$\rm erg\,s^{-1}\,cm^{-2}$}
\newcommand{\lcgs}{$\rm erg\,s^{-1}$}
\def\farcs{\hbox{$.\!\!^{\prime\prime}$}} 

\title[J1000+0234 with MUSE]{The  Ly$\alpha$, C{\sc iv}, and He{\sc ii} nebulae around J1000+0234: a galaxy pair at the center of a galaxy overdensity at $z=4.5$}

\author[E.F. Jim\'enez-Andrade et al.]{E.F. Jim\'enez-Andrade,$^{1, 2, 3}$\thanks{E-mail: ejimenez@irya.unam.mx} S. Cantalupo,$^{4}$ B. Magnelli,$^{5}$ E. Romano-Díaz,$^{3}$
\newauthor
C. G\'omez-Guijarro,$^{5}$
R. Mackenzie,$^{6}$
V. Smol\v{c}i\'c,$^{7}$
E. Murphy,$^{2}$
J. Matthee,$^{6}$ 
\newauthor
and
S. Toft,$^{8, 9}$\\
$^{1}$
Instituto de Radioastronomía y Astrofísica, Universidad Nacional Autónoma de México, Antigua Carretera a Pátzcuaro \# 8701,\\ Ex-Hda. San José de la Huerta, Morelia, Michoacán, México C.P. 58089\\
$^{2}$
National Radio Astronomy Observatory, 520 Edgemont Road, Charlottesville, VA 22903, USA\\
$^{3}$Argelander Institut f\"ur Astronomie, Universit\"at Bonn, Auf dem H\"ugel 71, Bonn, D-53121, Germany\\
$^{4}$ 
Dipartimento di Fisica G. Occhialini, Università degli Studi di Milano Bicocca, Piazza della Scienza 3, 20126 Milano, Italy\\
$^{5}$ Université Paris-Saclay, Université Paris Cité, CEA, CNRS, AIM, 91191 Gif-sur-Yvette, France\\
$^{6}$ Department of Physics, ETH Zurich, Wolfgang-Pauli-Strasse 27, CH-8093 Zurich, Switzerland\\
$^{7}$ Department of Physics, Faculty of Science, University of Zagreb, Bijeni\v{c}ka cesta 32, 10000 Zagreb, Croatia \\ 
$^{8}$ Cosmic Dawn Center (DAWN), Copenhagen, Denmark\\ 
$^{9}$ Niels Bohr Institute, University of Copenhagen, Jagtvej 128, DK-2200 Copenhagen, Denmark\\
}
\date{Accepted XXX. Received YYY; in original form ZZZ}

\pubyear{2022}

\begin{document}
\label{firstpage}
\pagerange{\pageref{firstpage}--\pageref{lastpage}}
\maketitle

\begin{abstract}
{  \Lya\,$\lambda$1216 (\Lya)} emission extending over $\gtrsim\,\rm 10\,kiloparsec\,(kpc)$ around dusty, massive starbursts at $z\gtrsim3$  might represent a short-lived phase in the evolution of present-day, massive quiescent galaxies. To obtain empirical constraints on this emerging scenario, we present  \Lya, {  \civ$\lambda$1550 (\civ)}, and {  \heii\,$\lambda$1640 (\heii)} observations taken with the Multi Unit Spectroscopic Explorer  towards   J1000$+$0234: a galaxy pair at $z=4.5$ composed of a low-mass starburst (J1000$+$0234$-$South) neighboring  a massive Submillimeter Galaxy (SMG; J1000$+$0234$-$North) that harbors a rotationally supported gas disk. Based on the spatial distribution and relative strength of \Lya, \civ, and \heii,  we find that star formation in J1000+0234$-$South and an active galactic nucleus in J1000+0234$-$North are  dominant factors in driving the observed 40\,kiloparsec-scale  Ly$\alpha$ blob (LAB). We use the non-resonant \heii\,line to infer kinematic information of the LAB. We find   marginal evidence for two spatially and spectrally separated  \heii\, regions, which suggests that the two-peaked \Lya\, profile is mainly a result of two overlapping {  and likely interacting} H{\sc i} clouds.  We also report the serendipitous identification of three Ly$\alpha$ emitters  spanning over a redshift bin $\Delta z \leq 0.007$ (i.e., $\lesssim 380\,\rm km\,s^{-1}$) located at  $\lesssim 140\,\rm kpc$ from J1000+0234. A galaxy overdensity analysis confirms  that J1000+0234 lies near the center of a Megaparsec-scale galaxy overdensity at $z= 4.5$ that might evolve into a galaxy cluster at $z=0$. The properties of J1000+0234 and its large-scale environment strengthen the link between SMGs within LABs, tracing overdense regions, as the progenitors of local massive ellipticals in galaxy clusters.
	\end{abstract}

\begin{keywords}
galaxies: high-redshift -- galaxies: evolution -- galaxies: intergalactic medium -- galaxies: individual: J1000+0234
\end{keywords}



\section{Introduction}

Extended nebulae of hydrogen {  \Lya\,$\lambda$1216 (\Lya) } emission  have become  key observational signatures of 
high redshifts structures  \citep[see][for a review]{ouchi20}. The {  most extended, with scale lengths}  $\gtrsim 100\,\rm kiloparsec\,(kpc)$, are associated with quasi-stellar objects \citep[QSOs, e.g.,][]{steidel00,  cantalupo14, cantalupo19} and  reside in overdense environments \citep[e.g.,][]{hennawi15, herez20, qiong21}. The most compact nebulae,  with {  scale lengths ranging  from 1$-$20\,kpc}, have been found around star-forming galaxies (SFGs) at $z\approx 2-7$ via stacking \citep[e.g.,][]{steidel11, matsuda12, xue17} and even direct detections
\citep[e.g.,][]{wisotzki16,  leclercq17, leclercq20}, suggesting that Ly$\alpha$ halos (LAHs) are ubiquitous in high-redshift galaxies \citep{wisotzki18}. There is an intermediate population of Ly$\alpha$ nebulae with {  scale lengths} of $10-100$\,kpc and Ly$\alpha$ luminosity
of $\sim 10^{43}\,\rm erg\, s^{-1}$ that are usually labelled as Ly$\alpha$ blobs \citep[LABs, e.g.,][]{matsuda04,shibuya18}. These LABs are associated with a wide range of galaxy populations, including  SFGs \citep[e.g.,][]{caminha16, geach16b}, high-redshift radio galaxies \citep[e.g.,][]{mccarthy87,swinbank15, marques-chaves19, wang21}, quasars/QSOs \citep[e.g.,][]{borisova16, farina17, ginolfi18, travascio20}. LABs also tend to lie in galaxy overdensities \citep[e.g.,][]{matsuda04, erb11,  alexander16, badescu17} and, as a result, these extended nebulae can be associated with multiple galaxies \citep{geach14, geach16b, guaita22, solimano22}.

Particular emphasis is given to  $z\gtrsim3$ LABs  surrounding  dusty, highly active SFGs selected at sub-mm wavelengths, typically known as  Submillimeter Galaxies \citep[SMGs, e.g., ][]{geach05, geach14, geach16b,  hine16, guaita22}. The  rare population of LABs around dusty galaxies with infrared luminosities $\log(L_{\rm IR}/L_\odot)\gtrsim 12$  has an average source density of only  $\sim 0.1\,\rm deg^{-2}$ \citep{bridge13}, likely because these systems  undergo a short-lived,  intense feedback phase that transforms  high-redshift starbursts into  mature/quenched systems \citep[e.g., ][]{bridge13, toft14}. LABs around luminous SMGs in rich environments might thus represent an early stage in the assembly of  $z\gtrsim2$ massive, quiescent  galaxies   \citep[e.g., ][]{gobat12, glazebrook17} and  local massive passive elliptical galaxies \citep[e.g.,][]{toft14, stach21}. To explore these scenarios, \emph{ an accounting of energetic processes and environment around SMGs within LABs is needed}. Nevertheless, information on  the ionizing photon budget and overall kinematics of LABs is hampered by the resonant nature of \Lya\, line emission and, in the specific case of SMGs,  dust absorption.

First, extended \Lya\, emission  can be a result of resonant scattering of \Lya\, photons  associated with star formation \citep[e.g., ][]{dijkstra09, behrens14,faucher-giguere15,  lake15, momose16,  mas-ribas17},   gravitational cooling radiation \citep[e.g., ][]{fardal01, dijkstra09b}, {  Ly$\alpha$ fluorescence, i.e., recombination radiation produced by gas photo-ionized by active galactic nuclei  (AGN) and/or  cosmic ultraviolet (UV) background \citep[e.g.,][for a review]{cantalupo17}}, and shock-heated gas by galactic superwinds \citep[e.g., ][]{taniguchi00}.
Thus, complementary to \Lya\, observations, knowledge of other UV emission lines such as {  \civ$\lambda$1550 (\civ)} and {  \heii\,$\lambda$1640 (\heii)} are necessary to infer better the physical processes powering \Lya\, emission \citep[e.g.,][]{villar-martin97, villar-martin07b, cantalupo19}. The C{\sc iv}/Ly$\alpha$ and He\,{\sc ii}/Ly$\alpha$ line ratios, in particular, can be compared with photoionization model predictions to disentangle the  contribution of stellar-photoionized and AGN-ionized gas to the observed \Lya\, line emission
\citep{villar-martin07b,arrigoni15, caminha16, cantalupo19, humphrey19,  marques-chaves19, herez20}.
Additionally, \civ\, line observations alone can inform about the distribution of  metal-enriched gas in high-redshift nebulae \citep[e.g.,][]{marques-chaves19}.

Second, despite the significant progress made to reproduce the observed \Lya\, properties of  galaxies at high redshifts via radiative transfer modeling \citep[e.g.,][]{zheng10, sadoun19, song20, gurung-lopez21},  kinematic information inferred from \Lya\, emission  \citep[e.g.,][]{verhamme06, orsi12} is highly model-dependant. Given the non-resonant nature of  \heii, this line arises as an unbiased probe of the gas kinematics \citep[e.g.,][]{cantalupo19}. Unfortunately, observations of the relative faint \heii\, and \civ\, lines in high-redshift galaxies are sparse \citep{prescott09, caminha16}, and have been preferentially detected in extreme environments around bright AGN/QSO \citep{dey05, cai17, cantalupo19, marino19, marques-chaves19, denbrok20}.

 Here, we use the Multi Unit Spectroscopic Explorer (MUSE) instrument to detect and characterize extended  \Lya, \civ, and \heii\, emission around  J1000$+$0234: a pair of galaxies at $z=4.5$ constituted by a {     SFG  with low stellar mass} ($\log(M_\star/M_\odot)=9.2\pm 0.1$)  neighboring a massive ($\log(M_\star/M_\odot)=10.1\pm0.1$) SMG \citep{gomez-guijarro18}.  We combine these observations with existing data from the Atacama Large Millimeter Array (ALMA) and  {\it Hubble Space Telescope} ({\it HST}) to study the ionizing mechanisms, kinematics, and large-scale environment of the LAB around J1000+0234. We use this knowledge to evaluate  evolutionary links between luminous SMGs in rich environments at $z>3$, LABs, and  quiescent systems in the center of present-day galaxy clusters.  This manuscript is organized as follows.  The properties of J1000+0234 are given in  Section \ref{sec:j1000}. The details of the observations and data analysis are presented in Section \ref{sec:observations},  while the results are shown in Section \ref{sec:results}. Finally, we discuss and summarize our main results in Section \ref{sec:discussion}. Throughout, we assume a cosmology of $h_0 = 0.7$, $\Omega_M = 0.3$, and $\Omega_\Lambda = 0.7$.

\begin{figure*}
	\begin{center}
		\includegraphics[width=17.6cm]{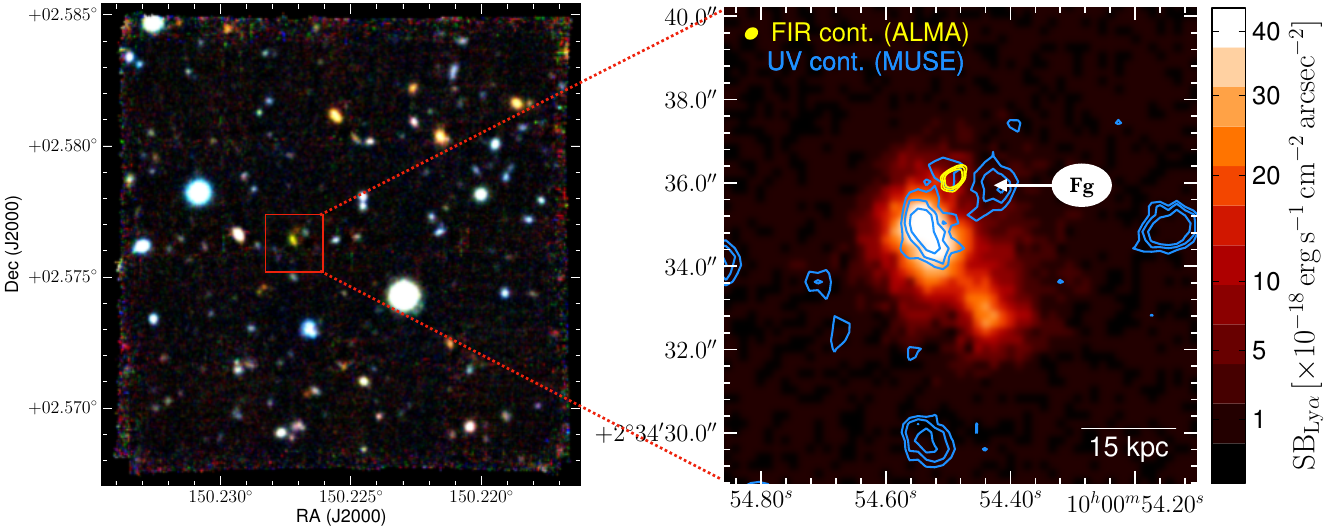}
		\caption{\textit{Left panel:} Three-color image of the MUSE data cube obtained towards J1000+0234. The pseudobands used to create this RGB image have been defined as ``blue" (4875\AA-6125\AA), ``green'' (6125\AA-7375\AA), and  "red" (7375\AA-8625\AA). The red square illustrates the zoomed-in region shown in the  \textit{right panel}, which displays the optimally extracted surface brightness (SB) maps of  detected \Lya\, emission of J1000+0234. The blue contours show the rest-frame UV continuum emission detected by MUSE (at $\sim 1200$\,\AA\, for $z=4.5$), while the {  yellow} contours illustrate the rest-frame far-infrared (FIR) continuum emission revealed by ALMA \citep[at $\rm \sim160\,\mu m$ for $z=4.5$;][]{gomez-guijarro18}. {  The  ALMA beam shape is shown in the top-left corner.}   The  UV and FIR emission contours are at 3, 5, {  and 8}  times  the rms noise level.  The thin white contour indicates the SNR = 3 isophote of the \Lya\, SB map.  The locus of the foreground "Fg"  galaxy \citep[$z=1.41;$][]{capak08} exhibiting UV continuum emission is indicated by the white arrow.  }
		\label{fig:muse_rgb_Lya}
	\end{center}
\end{figure*}

\section{J1000+0234}\label{sec:j1000}
	
The system J1000+0234 was {  identified in the 1.1\,mm map of the COSMOS field obtained with the AzTEC camera  \citep{wilson08} and the James Clerk Maxwell Telescope \citep{scott08}. J1000+0234 was subsequently associated with
a} Lyman Break Galaxy (LBG)  using long-slit spectroscopy with the DEIMOS instrument at the Keck telescope, which revealed a prominent double-peaked \Lya\, line profile  at $z=4.547$ \citep{capak08}.  
$^{12}\textrm{CO}(2-1)$ and $^{12}\textrm{CO}(4-3)$ line observations unveiled a  massive cold gas reservoir of $M_{\textrm{ gas}}\simeq2.6\times10^{10}M_\odot$ \citep{schinnerer08} in J1000+0234.
Imaging with the {\it  HST} and the Wide Field Camera~3 (WFC3), tracing the rest-frame UV emission,  revealed that J1000+0234  is composed of two major --likely interacting-- components separated by a projected distance of $\approx 6\textrm{\,kpc}$  \citep[see Figure\,\ref{fig:muse_rgb_Lya}; ][]{gomez-guijarro18}. The northern and more massive component ($\log(M_\star/M_\odot)=10.1\pm0.1$; J1000+0234$-$North) is embedded in a dusty environment with \hbox{$\log(L_{\textrm{IR}}/L_\odot)= 12.6\pm0.6$} and total star formation rate $\textrm{ SFR}=500^{+1200}_{-320}\,M_\odot \, \textrm{yr}^{-1}$ \citep{gomez-guijarro18}.
A strong, double-peaked [C\,{\sc ii}]\,158\,$\mu$m emission line was detected with  ALMA at the position of J1000+0234$-$North  (at $z=4.540$), indicating that this is a single, rapidly rotating disk with a dynamical mass of $\log(M_{\textrm{dyn}}/M_\odot)=11.4-11.6$ \,\citep{jones17, fraternali21}. 

The southern component, J1000+0234$-$South,  has a stellar mass of $\log(M_\star/M_\odot)=9.2\pm0.1$, thus $\sim$10 times lower than	that of J1000+0234$-$North. Yet, it emits the bulk of the rest-frame UV emission ($75$\,percent) observed with \textit{HST}, consistent with the fact that no dust continuum (nor [C{\sc ii}] line) emission has been detected towards this component \citep{capak08,gomez-guijarro18}. The UV emission indicates  a   ${\rm SFR}=148\pm8\, M_\odot\, \textrm{yr} ^{-1}$ and a UV slope ($\beta$) of $-2$,  as expected for LBGs at similar redshift \citep[][and references therein]{gomez-guijarro18}. Finally,  the SFR and stellar mass of J1000+0234$-$South and J1000+0234$-$North imply that these galaxies lie at 1.3 and 1.0\,dex, respectively, above the main sequence of SFGs at $z=4.5$ \citep[assuming the relation from][]{schreiber15}. In other words, the SFRs of J1000+0234$-$North and J1000+0234$-$South  are one order of magnitude higher than that of normal, main-sequence galaxies at the same redshift and stellar mass range. We, therefore, deem J1000+0234$-$South and J1000+0234$-$North as  starburst galaxies.\\

\section{MUSE observations and data analysis}\label{sec:observations}

\subsection{Observational data and reduction}
J1000+0234 has been observed with MUSE mounted on the Very Large Telescope (VLT-Yepun, UT4) using 16 exposures of 900\,s {  and ground-layer adaptive optics. Each exposure was dithered slightly (randomly by $\approx$1-2\,arcsec) and rotated by 90 degrees compared to the previous exposure. These data were obtained over three observing blocks on March 7, 2019, April 8, 2019, and May 5, 2019,} under good weather conditions with an average seeing $\rm FWHM\approx0.9\rm\, arcsec$ and airmass $<1.4$. These observations, with a total integration time on source of 4\,hrs, are part of the ESO GTO Programs 0102.A-0448 and 0103.A-0272 (PIs: S. Lilly and  S. Cantalupo). The {  standard reduction steps (bias subtraction, flat-fielding, wavelength and flux calibration, geometrical cube reconstruction) are} performed using  the ESO MUSE pipeline \citep[{  version 2.6;}][]{weilbacher14}   without performing sky-subtraction. Then, we employ the {\tt CubExtractor} package [{\tt CubEx} hereafter, see \citet{cantalupo19} for a description] to improve flat-fielding and perform sky subtraction using  {\tt CubeFix} and {\tt CubeSharp}. {  Such an additional flat-fielding with {\tt CubeFix} is implemented to reduce significant residuals in the  white-light images generated by the pipeline \citep[see Section 2.1 of][for a detailed description of this step]{cantalupo19}}. We combine the individual exposures using {\tt CubeCombine} that applies an average-sigma clipping method. {  Half of the exposures  were  affected by  
intra-dome light contamination around $8000-9000$\,\AA\,  due to a malfunctioning monitoring camera\footnote{\url{https://www.eso.org/sci/facilities/paranal/instruments/muse/news.html}; see the entry from May 31, 2019.}. Hence,} we created two different data cubes with {\tt CubeCombine}. The first data cube only contains ``clean" exposures ("cleanexp" data cube hereafter) that account for 50\,percent of all the available data. It is, therefore,  suitable for identifying \civ\, and \heii\,   at the expected observed wavelength of $ 8560$\,\AA\,  and $9086$\,\AA\,, respectively. The second data sample, hereafter "allexp" cube, contains all the exposures, regardless if they are affected or not by the issue mentioned above.  The effective/clean spectral region of this data cube is  $5000-8000$\,\AA\,,  allowing us to probe \Lya\, emission at the expected observed wavelength of $6742$\,\AA\,.  {  The "cleanexp" and "allexp" data cubes have a $1\sigma$ noise level of} $\approx$2.0 and $1.4\,\times\rm10^{-20} erg\,s^{-1}\,cm^{-2}$\,arcsec$^{2}$, respectively,  per layer (at 6650\,\AA)\, within an aperture of 1\,arcsec$^{2}$. Both data cubes have a spectral resolution of 1.25\,\AA\, and pixel scale of 0.2\,arcsec.

\begin{figure*}
	\begin{center}
		\includegraphics[width=17.6cm]{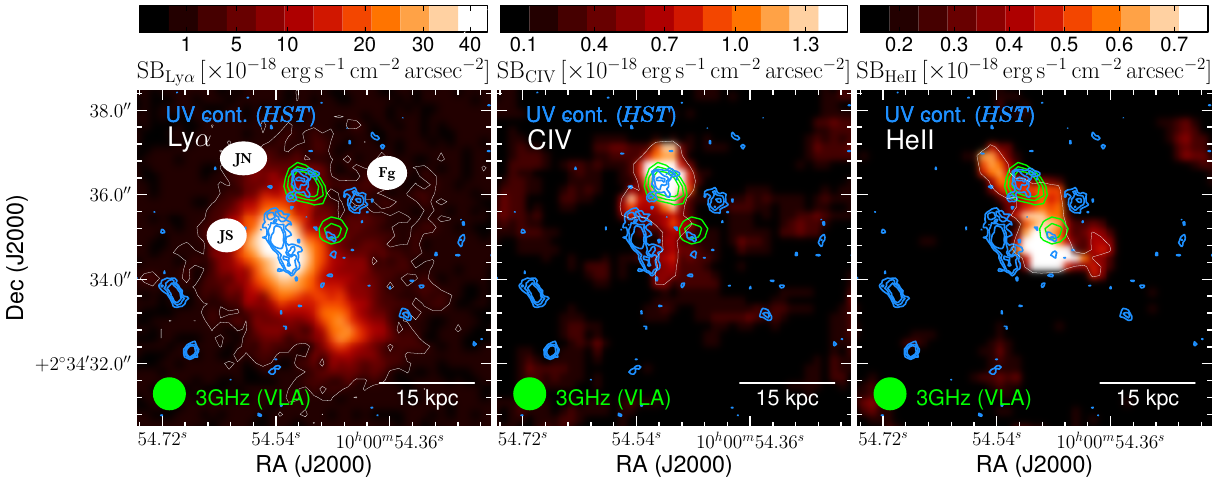}
		\caption{Optimally-extracted SB maps of  \Lya\, (\textit{left panel}), \civ\,  (\textit{middle panel}), and \heii\, (\textit{right panel})  emission of J1000+0234. In the case of \civ\, and \heii, we apply a spatial boxcar smoothing ($2\times2$ pixels) to the cube before producing the SB maps (using the option {\tt -boxsm} in {\tt CubEx}).  The thin white contours indicate the SNR = 3 isophote. 	The blue contours show the rest-frame UV continuum from the \textit{HST}/F160W imaging and the lime contours display the VLA 3\,GHz radio continuum emission at $0\farcs75$ resolution \citep{smolcic17}. {  The former are plotted at 3, 5, and 8  times the rms noise level, while the latter are drawn at 3, 4, and 5  times the rms noise level. The VLA beam shape is shown at the bottom-left corner.} There are three prominent UV  sources within the extended Ly$\alpha$ blob: a southern component associated with a low-mass SFG (J1000+0234$-$South: JS), a northern UV emitting source linked to a massive SMG (J1000+0234$-$North; JN) with a 3\,GHz radio counterpart, and a foreground ``Fg" galaxy at $z=1.41$ \citep{capak08, gomez-guijarro18}. While the brightest  \Lya\, region matches the locus of the low-mass SFG, the brightest \civ\, region is coincident with the SMG position. \heii\, emission {  is maximal at $\approx 0\farcs6$ to the South-East of a} compact 3\,GHz radio source that has a faint UV counterpart.    }
		\label{fig:vel-int_lines}
	\end{center}
\end{figure*}

\subsection{Continuum subtraction and three-dimensional signal extraction}
\label{subsec:contsub}

The relative astrometry of  MUSE data cubes  is  accurate ($\approx$0.05\,arcsec noise  rms),  
yet the "absolute"  world coordinate system (WCS)  can be off by a few arcsec. 
To calibrate the WCS of the ``allexp'' and ``cleanexp'' cubes, we use  the position of bright point-like sources in the  cubes and cross-match them with the coordinates inferred from {\it HST} imaging \citep{gomez-guijarro18}, which was aligned to the absolute astrometry of the COSMOS field ensuring an absolute astrometric accuracy of $\approx 0\farcs1$, or better, that is half of one pixel for our MUSE data cubes.  \\
We use the routine {\tt CubeBKGSub} from the {\tt CubEx} package to subtract the continuum of both cubes.  We first mask the following spectral regions that exhibit evidence of line emission: 6725--6800\AA\, (around \Lya), 8572--8612\AA\, (around \civ), and  9071--9111\AA\, (around \heii). Then, we apply median filtering along the spectral axis  using windows of 100 and 40 pixels for the \Lya, and \civ/\heii\, spectral cube, respectively.  We also smooth  the resulting continuum  across two neighboring spectral bins. \\
A sub-cube that contains the spectral region around \Lya\, emission is extracted from the  continuum-subtracted ``allexp'' cube. This  encompasses the layers/channels from 1580 to 1700, corresponding to the wavelength range of  6673.63--6823.63\,\AA.\, Similarly, two sub-cubes that cover the wavelength range around  \civ\, (channels 3050-3180;  8510.8--8672.06\,\AA)\, and \heii\, (channels 3495--3535;  9067.1--9117.1\,\AA)\, lines are extracted from the continuum-subtracted ``cleanexp" cube.  
We execute {\tt CubEx} on the subcube containing \Lya, line  emission using  the following configuration. We set {\tt RescaleVar= true} to re-scale the original variance in the cube using the variance estimated layer by layer. We adopt  {\tt FilterXYRad}=2  to apply a Gaussian filter along the spatial direction using a 2-pixels radius. No Gaussian filter is applied along the wavelength direction  (i.e., {\tt FilterZrad}=0). For the detection, we set a voxel individual threshold in signal-to-noise ratio ({\tt SN-Thereshold}) of 2.5 and a minimum number of voxels ({\tt MinNVox}) of 500. No minimum number of spatial/spectral pixels is adopted for detection (i.e., {\tt MinArea}=0 and {\tt MinDz}=0).  We also mask the noisy edges of the data cube using  {\tt XYedges}=30\,pixels ($6\farcs0$).   All the other parameters are set to the default values given by {\tt CubeEx}.  Finally, to optimize the three-dimensional detection of the faint and compact \civ\,and \heii\,  emission, we run  {\tt CubEx} using  {\tt MinNVox}=100. The parameters {\tt RescaleVar}, {\tt FilterXYRad}, {\tt FilterZrad}, {\tt SN-Thereshold}, {\tt MinArea}, and {\tt MinDz} are set to the same values used to detect \Lya\, emission.

\begin{table*}
	\caption{Properties of \Lya, \civ, and \heii\, emission lines of the full J1000+0234 system derived from our 3D line extraction procedure described in \S\,\ref{sec:results}. }
	
	\label{tab:line_properties}
	\begin{tabular}{lccccc}
		\hline
		\hline
		Emission line & RA & DEC &  Proj. area & Iso Flux & Line luminosity \\
		& [hh:mm:ss.sss] & [dd:mm:ss.ss]  &  [arcsec$^{2}$\,/ \, kpc$^{2}$] & [$\times 10^{-17}$\fcgs] &  [$\times 10^{42}$\lcgs]\\
		\hline
		Ly$\alpha$ & 10:00:54.510   & 02:34:34.27  &   42.9 / 1853 & $19.79\pm 0.15$ &  $41.17 \pm 0.31$   \\
		
		\civ & 10:00:54.509   & 02:34:35.94  &  3.7 / 159 & $0.37\pm 0.03$ &  $0.76 \pm 0.07$   \\
		
		\heii & 10:00:54.478   & 02:34:35.18  &    5.5 / 238 & $0.27\pm 0.03$ &  $0.57 \pm 0.06$    \\		 
		\hline 
	\end{tabular}
\end{table*}

\section{Results}\label{sec:results}

To explore the spatial distribution of  the \Lya, \civ,\, and \heii\,  emission lines,  we obtain  surface brightness (SB) maps  with the task {\tt cube2im} (see Figure\,\ref{fig:vel-int_lines}).   We use  the {\tt idcube} inferred by {\tt CubEx} while performing the 3D line extraction presented in \S\,\ref{subsec:contsub}. Such an {\tt idcube}  contains  the three-dimensional masks  associated with the source/line of interest. 
We also employ the task {\tt cube2spc} to derive the 1D spectra of the \Lya, \civ, and \heii\,    emission lines (see Figure\,\ref{fig:lines}). Because in deriving these 1D spectra we only consider the voxels in the three-dimensional masks, i.e., voxels with a sufficiently high signal-to-noise ratio ($\rm SNR >2.5$), we deem these spectra as ``optimally-extracted''. The key properties of the \Lya, \civ, and \heii\, emission lines reported in the {\tt CubEx} catalogs are presented in Table\,\ref{tab:line_properties}. These are discussed in more detail in the following paragraphs.

\subsection{Ly$\alpha$\, emission line}

\begin{figure}
	\begin{center}
		\includegraphics[width=6.5cm]{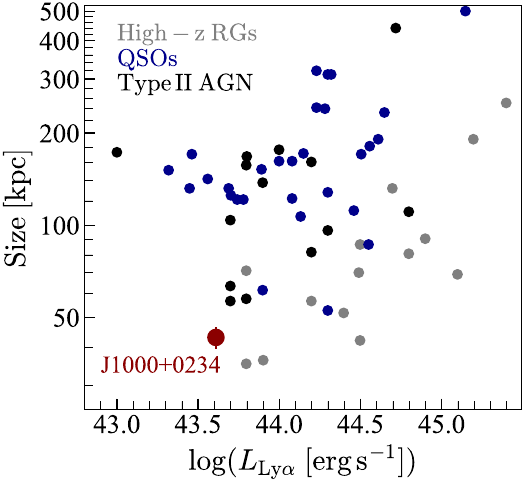}
		\caption{Total extent as a function of Ly$\alpha$ luminosity of LABs previously associated with high-redshift radio galaxies (High-z RGs), QSOs, and Type\,II AGN \citep[using data from][and references therein]{marques-chaves19, sanderson21, wang21}. The total extent of the LABs around J1000$+$0234 is inferred using $\sqrt{a}$, where $a$ is the projected area within which  \Lya\, is significantly detected with {\tt CubEx}. The Ly$\alpha$ nebulae around the J1000$+$0234 system is among the most compact and dimmest LABs at $z\gtrsim 2$.  }
		\label{fig:comparison_size-vs-llya}
	\end{center}
\end{figure}

In the left panel of Figure\,\ref{fig:vel-int_lines}, we overlay the rest-frame UV continuum emission from the  {\it HST}/F160W imaging \citep[with a $\approx0\farcs2$ resolution; ][]{gomez-guijarro18} over the  velocity-integrated \Lya\, line emission map of J1000+0234. This overlay indicates that the brightest \Lya\, emitting region  spatially correlates with the brightest UV component of the J1000+0234 complex. Interestingly, the rest-frame UV emission is elongated along the same direction of the \Lya\, nebula. This is further verified by the spatial distribution of the  rest-frame  UV continuum emission from the MUSE imaging (see the right panel of Figure \ref{fig:muse_rgb_Lya}).  This UV-bright, low-mass SFG that we identify as JS ($\equiv$J1000+0234$-$South) appears  to be linked to the ionizing source of the extended \Lya\, emission (see Section \ref{sec:discussion}). Our extraction procedure with {\tt CubEx} indicates that such a Ly$\alpha$ nebula extends over a projected area of 42.9\,arcsec$^2$ $\approx1853\,\rm kpc^{2}$.
With an {  observed} maximum linear projected size of  $\rm \approx43\,kpc$, the emission around J1000+0234 {  appears to be} among the most compact  and faintest Ly$\alpha$ nebulae detected thus far 
if compared to nebulae detected around other AGN and bright quasars (Figure\,\ref{fig:comparison_size-vs-llya}). {  Since the observed total extent strongly depends on the detection limit, we note that the reported maximum linear size should be strictly interpreted as a lower limit to the full extent of the Ly$\alpha$ nebula. This lower limit, however, suffices to deduce that the Ly$\alpha$ nebula of J1000$+$0234 is} brighter and larger than the typical Ly$\alpha$ halos detected around galaxies \citep[e.g.,][]{leclercq20}.

 \begin{figure*}
	\begin{center}
		\includegraphics[width=17.6cm]{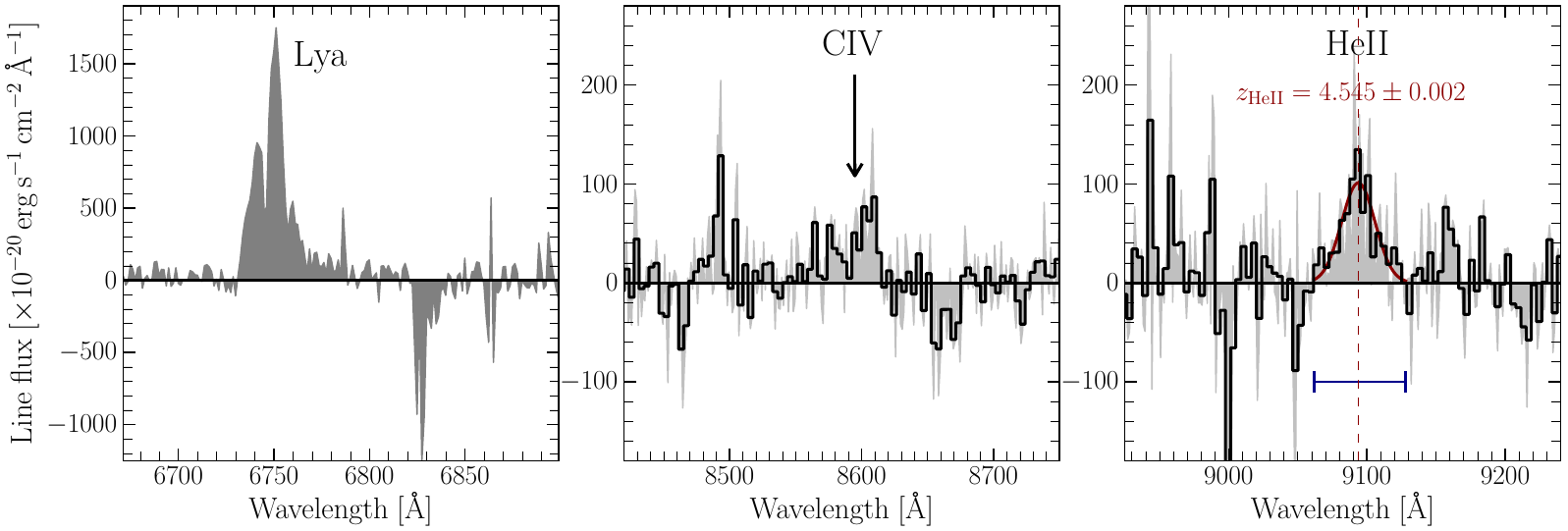}
		\caption{Optimally-extracted 1D spectrum of the \Lya\, (\textit{left panel}), \civ\,  (\textit{middle panel}), and \heii\, (\textit{right panel}) emission lines of J1000+0234. The solid histograms show the \civ\, and \heii\,  spectra averaged over a 4.5\AA-width bin. {   The strong negative features around 6830\,\AA\, are sky line residuals.} The red line shows a Gaussian model to fit the \heii\, line, and the vertical red-dashed line marks the central wavelength of the profile. The non-resonant \heii\,  line indicates a redshift of $4.545\pm0.002$ for the J1000+0234 system. The horizontal bar  shows the wavelength range used to fit the Gaussian model.}
		\label{fig:lines}
	\end{center}
\end{figure*}

The observed (optimally-extracted) flux of $F_{\rm Ly \alpha}^{\rm obs}=19.79\pm 0.15 \times 10^{-17}$\fcgs\, leads to a total line luminosity of  $F_{\rm Ly \alpha}^{\rm obs}=40.17\pm 0.31 \times 10^{42}$\lcgs, assuming $z=4.545\pm0.0002$ (see the following Section), which sets J1000+0234  at the faint end of the luminosity distribution of Ly$\alpha$ nebulae around  QSO, Type II AGNs, and high-redshift radio galaxies (Figure\,\ref{fig:comparison_size-vs-llya}).
The 1D optimally-extracted spectrum of J1000$+$0234 (left panel of Figure\,\ref{fig:lines}) reveals a prominent emission line around 6750\AA\, associated with Ly$\alpha$, which was originally reported by \citet{capak08}. This broad and asymmetric line 
exhibits a double peak, suggesting  complex gas kinematics in the Ly$\alpha$ nebula. We analyze and discuss the \Lya\, line profile in more detail in Section \ref{subsec:kinematics}.

\subsection{\civ\, and \heii\, emission line}

In the middle and right panel of Figure\,\ref{fig:vel-int_lines}, we present the optimally-extracted SB maps of  \civ\, and \heii\, line emission.  
While \civ\,  line emission peaks at the locus of J1000+0234$-$North (a massive, dusty starburst), the brightest \heii\, region {  encloses} a faint UV continuum source that is offset by $\approx1$\,arcsec to the west of the  low-mass SFG J1000+0234$-$South. Such a faint source, originally identified by  \citet{gomez-guijarro18}, is  visible in the {\it HST}/F814W,  {\it HST}/F125W, and {\it HST}/F160W imaging and has been labelled as a minor companion of the massive SMG J1000+0234$-$North \citep{gomez-guijarro18}.  There also exists a 3\,GHz radio source detected at the 4.5$\sigma$ level that lies at $<0\farcs5\,\rm arcsec$ from this faint UV component and has a flux density of $1.0\pm0.2\mu\rm Jy$ \citep[see our Figure\,\ref{fig:vel-int_lines};][]{smolcic17}. 

Our source extraction procedure with {\tt CubEx} indicates that \civ\, and \heii\, emission  extends out to a projected area of $\rm 3.7\,arcsec^2\approx159\,\rm kpc^2$ and $\rm 5.5\,arcsec^2\approx238\,\rm kpc^2$, respectively. With a total flux (i.e., flux of all voxels within the 3D detection mask) of $F_{\rm CIV}^{\rm obs}=0.37\pm0.03 \times 10^{-17}$\,\fcgs\, and  $F_{\rm HeII} ^{\rm obs}=0.27\pm0.03\times 10^{-17}$\,\fcgs, we infer total line luminosities of $L_{\rm CIV}=0.76\pm0.07\times 10^{42}$\,\lcgs\, and  $L_{\rm HeII}=0.57\pm0.06\times 10^{42}$\,\lcgs (see Table\,\ref{tab:line_properties}).  The optimally-extracted  spectra of the \civ\, and \heii\,  emission lines are shown in the middle and right panels of Figure\,\ref{fig:lines}. 
 Given the resonant nature of the \Lya\, and \civ\, emission lines, we use the  \heii\, line to infer the redshift of  the J1000$+$0234 system. 
 Fitting a 1D Gaussian model  we find a central wavelength of  $9093.8\pm1.9$\AA\, (and $\rm FWHM=12.1\pm1.9$\AA$\;\approx940\pm110\,$\kmps), which implies a  redshift of $z=4.545\pm0.002$ that is  comparable with that of J1000+0234$-$North  ($4.5391 \pm 0.0004$) derived from [C{\sc ii}] observations \citep{fraternali21}.
 
 \begin{figure*}
 	\begin{center}
 		\includegraphics[width=17.5cm]{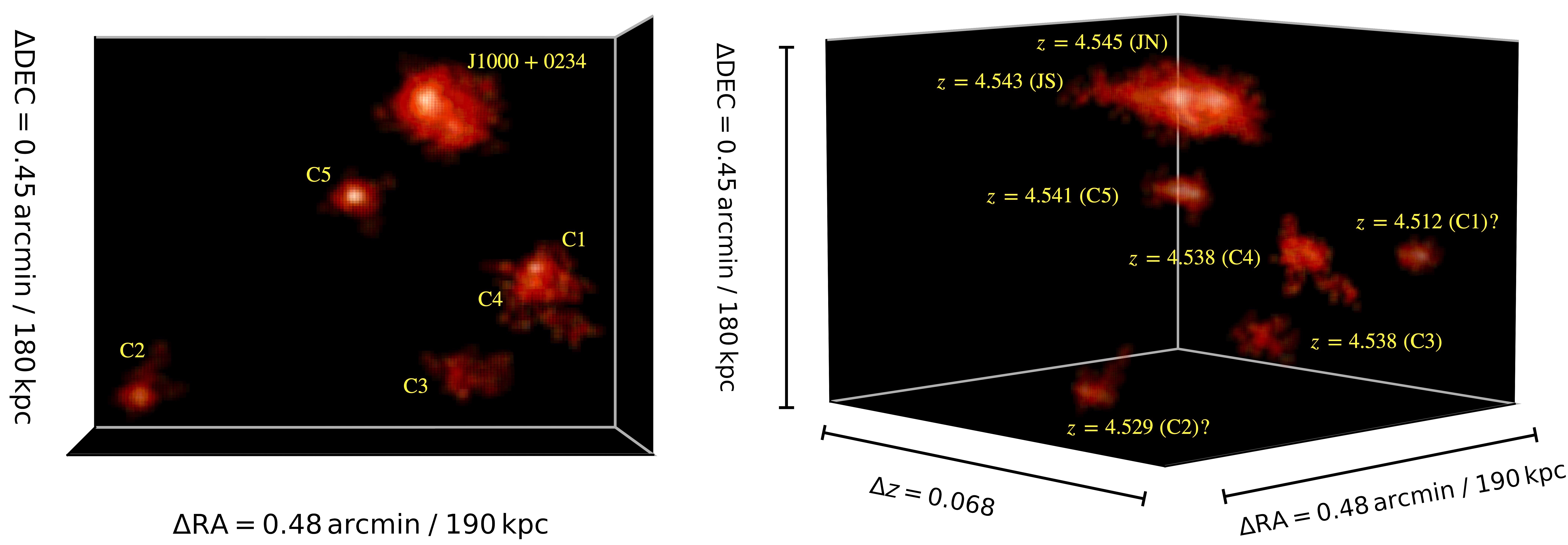}
 		\caption{\Lya\, line emission of J1000+0234 and five emission-line sources (C1-C5) identified within a $48\times48\rm \, arcsec^{2}$ area around J1000+0234: as observed on the plane of the sky ({\it left panel}) and along the wavelength/redshift axis ({\it right panel}). The color scale shows the flux density per voxel, normalized to the minimum and maximum value observed in this subcube.  The redshift values presented for the five companions are derived under the assumption that the emission line corresponds to \Lya.\,  The physical distance and redshift bins shown in the axis labels are computed at $z=4.545$. The discovery of these companions around J1000+0234 {  support earlier results suggesting that this system resides in} a galaxy overdensity, i.e., a protocluster, at $z=4.5$ \citep{smolcic17c}.  }
 		\label{fig:environment_2}
 	\end{center}
 \end{figure*}

\subsection{Large-scale environment of J1000+0234}\label{subsect:environment}

\begin{figure*}
    \begin{center}
    \includegraphics[width=17.5cm]{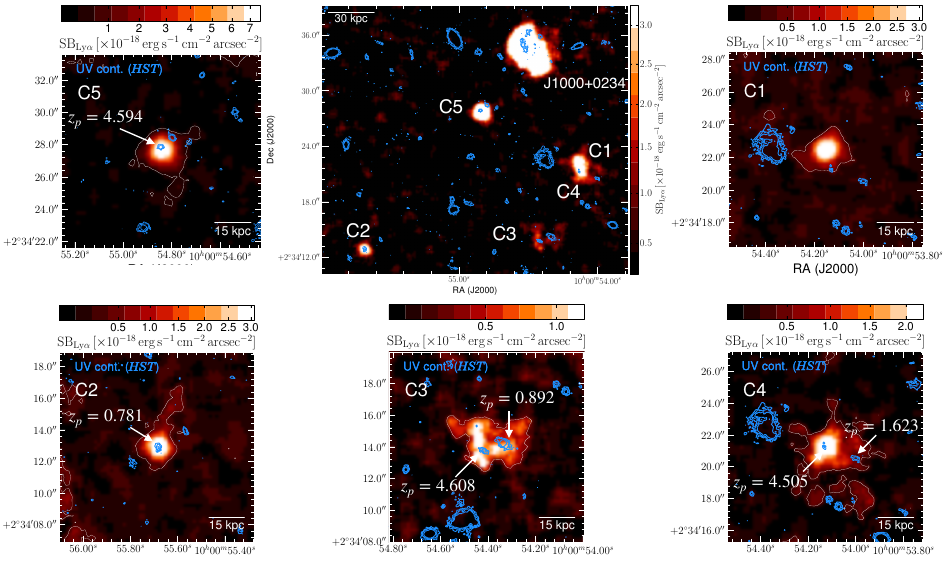}
 		\caption{Spatial distribution ({\it central panel}) and optimally extracted SB maps ({\it lateral/lower panels}) of the five emission-line sources in the vicinity of J1000+0234. {   The blue contours show the rest-frame UV continuum from the \textit{HST}/F160W imaging and are plotted at 3, 5, and 8  times the rms noise level.} The photometric redshifts (if available) of the UV continuum sources located within the extended emission-line regions are shown in white. The UV counterparts of C3, C4, and C5  have a photometric redshift of $z_p\approx 4.5$. Therefore, we can robustly associate the C3, C4, and C5 emission-line regions with \Lya.\, Because there is no UV counterpart for C1 and because C2 might correspond to  the [O{\sc ii}]\,$\lambda$3727 doublet of a  $z=0.803\pm0.001$ foreground galaxy,  we deem C1 and C2 as potential \Lya\, emitting sources at $z\approx4.5$.}  
        \label{fig:environment}
   \end{center}
\end{figure*}

\begin{figure}
 	\begin{center}
 		\includegraphics[width=7.5cm]{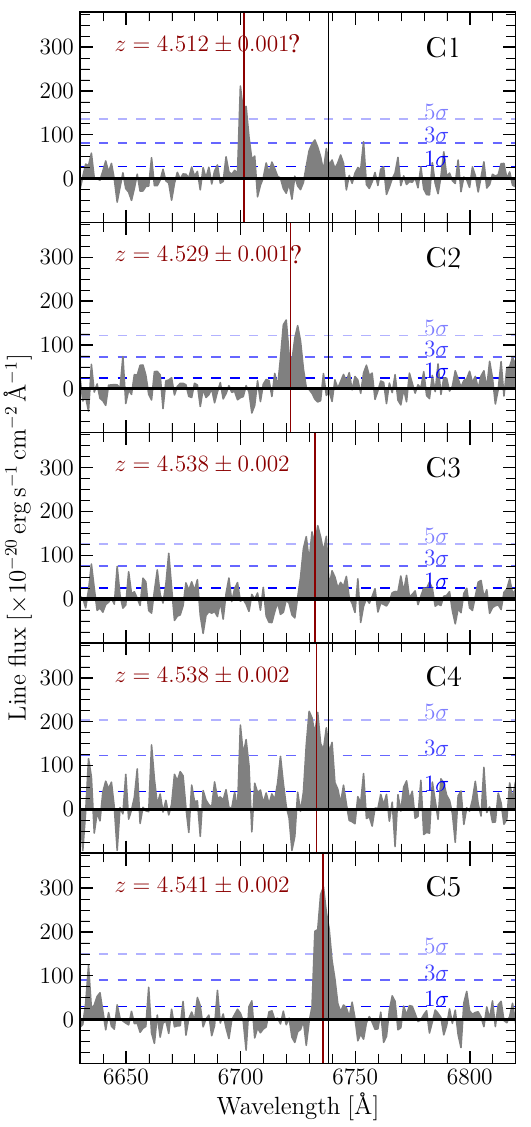}
 		\caption{Optimally-extracted 1D spectra of the five emitting sources in the vicinity of J1000+0234. The horizontal blue lines show the 1, 3, and 5 $\times$ rms noise level. The vertical red lines show the (flux-weighted) central wavelength of the  line profiles.  The vertical black line marks the observed wavelength of \Lya\, line emission of J1000+0234. The photometric redshifts of  C3, C4, and C5  allow us to  identify the observed lines at $\approx$6735\,\AA\,  as \Lya\, at $z\approx 4.5$. The properties of C1 and C2, and their UV counterparts, prevent a robust association between the observed $\approx$6710\,\AA\, lines with \Lya\, (see Section\,\ref{subsect:environment}).  }
 		\label{fig:environment_1dspecs}
 	\end{center}
 \end{figure}

The source extraction performed with {\tt CubEx} leads to the identification of five potential \Lya\, emitting sources  within the $48\farcs0\times48\farcs0$ masked FoV (due to noisy edges) of MUSE (Figure\,\ref{fig:environment_2}). The emission peaks of these sources are detected within a spectral window  6699.9--6743.6\AA\,,  implying a redshift offset $\delta z <0.033$ (and velocity offset  $\delta v < 1800\rm\,km\,s^{-1}$) with respect to J1000+0234. Extending the search for \Lya\, companions to larger spectral windows only reveals a  detection of a potential \Lya\, emitter  at $z=4.428$, which is too distant ($\delta z=0.117$, i.e., $\rm \delta v\approx 6300\,km\,s^{-1}$) from J1000+0234 to suggest membership. The  optimally-extracted SB maps of the five sources is presented in  Figure\,\ref{fig:environment}, while their optimally-extracted 1D spectra and inferred line properties are shown in Figure\,\ref{fig:environment_1dspecs} and Table\,\ref{tab:companions}, respectively. 

To verify that the observed line emission from the five neighboring sources around J1000+0234 corresponds to \Lya\, at $z\approx 4.5$, we identify their UV counterparts in the {\it HST}/WFC3 continuum imaging \citep{gomez-guijarro18} and crossmatch their positions with the CLASSIC COSMOS2020 photometric catalog \citep{weaver21}. Using a $1\farcs0$ search radius, we find the following: 

\begin{enumerate}
\item The two UV sources near the brightest emitting region of C5 are linked to a COSMOS2020 source with photometric redshift $z_p=4.594\pm0.050$, $\log(M_\star/M_\odot)=10.24^{+0.10}_{-0.13}$, and $\rm{SFR}=56^{+19}_{-11}$\,\mpyr.

 \item The central UV source in C4 has a photometric redshift  $z_p=4.505^{+0.330}_{-0.280}$, $\log(M_\star/M_\odot)=8.62^{+0.18}_{-0.22}$ and ${\rm SFR}=2.0^{+2.7}_{-0.5}$\,\mpyr. The UV source at the South-West of the central region of C4 is a foreground galaxy at $z_p=1.623^{+0.470}_{-0.290}$. 
 
\item  The more compact UV source in C3 lies at $z=4.608^{+0.190}_{-0.170}$, it has a stellar mass  $\log(M_\star/M_\odot)=9.79^{+0.15}_{-0.23}$, and ${\rm SFR}=25^{+44}_{-6}$\,\mpyr. The more extended UV source is a foreground galaxy at $z=0.892^{+0.050}_{-0.040}$. 

 \item The central UV source in C2 has a photometric redshift of $z_p=0.781\pm 0.050$, $\log(M_\star/M_\odot)=8.05\pm0.09$, and ${\rm SFR}=0.28^{+0.10}_{-0.12}$\,\mpyr. Based on this redshift estimate, the emission line of C2 at 6721.8\,\AA\, could  correspond to the [O{\sc ii}]\,$\lambda$3727 doublet at $z=0.803\pm0.001$.  Indeed, the inferred [O{\sc ii}]-based SFR \citep{kennicutt98} of $0.35\pm 0.10$\,\mpyr matches  that reported in the CLASSIC COSMOS2020 catalog.  {  Assuming  $z=0.803\pm0.001$},  the rest-frame separation between the peaks {  would be} $\approx 3$\AA\,, {  which} agrees with that of the [O{\sc ii}]\,$\lambda$3727 doublet. {  In addition, the symmetric double-peaked  line at 6721.8\,\AA\, would be at odds with the typical  Ly$\alpha$ line profiles of high-redshift galaxies where the blue peak can be significantly absorbed by the foreground intergalactic medium \citep[e.g.,][]{laursen11, hayes21}, although prominent blue peaks due to gas accretion are expected as well \citep{ao20}. Since} no other emission line is robustly detected at the locus of C2 in the MUSE data cube, {  we can not}  firmly associate the 6721.8\,\AA\, line with [O{\sc ii}]\,$\lambda$3727   and confirm that C2 lies at $z\approx0.8$.  Given the ambiguity in the nature of the 6721.8\,\AA\, line, we do not rule out C2 as a likely \Lya\, companion around J1000+0234. 
 
 \item No UV continuum counterpart is identified for C1. We deem this source as another potential \Lya\, emitter at $z=4.512$.
 
\end{enumerate}
 
In summary, the photometric redshifts of the UV counterparts of the extended C3, C4, and C5 sources support our initial assumption that their emission lines at $\approx$6735\,\AA\,  correspond to \Lya.\, Focusing on these three --robustly identified-- companions around J1000+0234, we observe that their \Lya\, luminosity ranges from 1.8 to $3.6\times 10^{42}$\lcgs. Interestingly, the luminosity of these sources and their redshift/velocity offset with respect to J1000+0234 are anti-correlated (Table\,\ref{tab:companions}), i.e.,  the more luminous companions are closer ({  both on the  plane of the sky and} along the line of sight) to J1000+0234. The nearest companion, C5, lies at a projected distance of $\approx50$\,kpc to the South-East of J1000+0234, while the farthest one (C3) lies at $\approx140$\,kpc to the South of J1000+0234. {  Whereas} \Lya\, emission in C5 is concentrated in a single component with a projected area of 315$-$460\,kpc$^{2}$, the \Lya\, emission of C3 and C4  displays an extended, clumpy morphology that spreads out to a projected distance of 600 and 1090\,kpc$^{2}$, respectively.

\begin{figure*}
 	\begin{center}
 		\includegraphics[width=15cm]{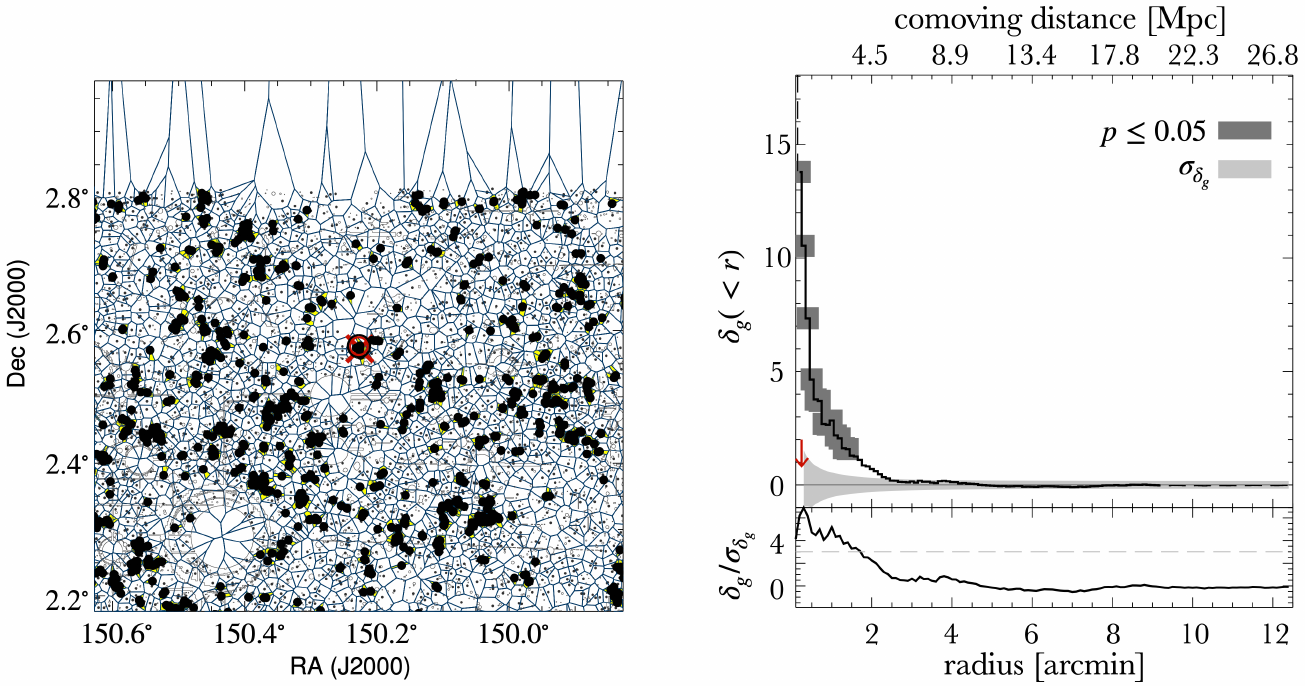}
 		\caption{ ({\it Left:}) Voronoi tessellation map around J1000+0234 (indicated by the red circle). Overdense VTA cells are shown in yellow.  The overdensity center is denoted by the red cross. ({\it Right:}) Galaxy overdensity as a function of distance to the overdensity center. The grey region shows the error on the galaxy overdensity parameter. The gray squares mark robust overdensity values, for which the Poisson probability of observing more or an equal number of sources than expected from the  background galaxy surface density is $\leq 5\%$.  The  projected distance from the overdensity center to J1000$+$0234 is indicated by the red downwards pointing arrow.  The bottom panel presents the significance of the overdensity $(\delta_{\rm gal}/\sigma_{\delta_{\rm gal}})$ as a function of radius. The value of $\delta_{\rm gal}/\sigma_{\delta_{\rm gal}}=3$ is indicated by the horizontal dashed line. }
 		\label{fig:environment_vta}
 	\end{center}
 \end{figure*}

\subsubsection{A galaxy overdensity around J1000+0234}

Using  the COSMOS2015 photometric catalog \citep{laigle16}, \citet{smolcic17c} anticipated that J1000+0234 belongs to a galaxy overdensity. This result is reinforced by the discovery of \Lya\, emitters at $z\approx 4.5$ in the $317\times317\,\rm kpc^{2}$ (unflagged) FoV of our MUSE data cube.
Here, we verify the overdensity by repeating the analysis presented by \citet{smolcic17c} using the new COSMOS2020 photometric redshift catalog \citep{weaver21} and the Ly$\alpha$-detected sources in this work. 

For this analysis, we restrict the COSMOS2020 CLASSIC catalog to (Subaru Suprime-Cam) sources with $i \leq 27.03$ to match the magnitude limit to our faintest MUSE source. The selected sources lie within the Ultravista area ($\mathrm{FLAG}\_\mathrm{UVISTA} = 0$) and fall within the redshift range of $\Delta z = 0.53$ centered at the redshift of J1000+0234.  No sources in the photometrically flagged areas in the Subaru Suprime-Cam data ($\mathrm{FLAG}\_\mathrm{SUPCAM} = 0$) are considered. The above criteria select J1000$+$0234$-$North, J1000$+$0234$-$South, and their three robustly identified companions (C3, C4, and C5) from the COSMOS2020 catalog.

We first run the Voronoi tessellation analysis which identifies overdensities by partitioning a plane into convex polygons called Voronoi cells. Each cell encloses exactly one point (in our case one galaxy in the chosen $\Delta z$ range). The inverse of the { volume} of a galaxy’s cell effectively corresponds to the local number density of galaxies for the given galaxy. The threshold above which a cell is considered “overdense” was chosen to be the 80\% quantile of the cumulative distribution of cell densities obtained using 10 Mock catalogs. The Mock catalogs were generated to contain the same number of galaxies as present in the inner 1 deg$^2$ of the COSMOS field, but with randomly distributed positions over this area. For more details about the method see Section\,3.1.2. in \citet{smolcic17c}. The Voronoi tessellation analysis results are presented in the left panel of Figure~\ref{fig:environment_vta}, where we indicate only galaxies occupying "overdense" cells. An overdensity, centered on the four sources, J1000$+$0234, C3, and C5, is clearly discernible.

We further compute the galaxy overdensity parameter ($\delta_g$) as a function of distance from the identified center of the overdensity that is RA$=$150.22578\,deg and DEC$=$2.572560\,deg. This is done by equaling the overdensity center to the average right ascension and declination of sources within the red circle in the left panel of Figure\,\ref{fig:environment_vta} (see Section\,3.1. in \citet{smolcic17c} for more details).  The galaxy overdensity parameter, which is the contrast above the background field, is defined as a function of radius ($r$):
\begin{equation}
\label{eq:delta}
\delta_{\rm g}(r)\equiv \frac{\Sigma_{\rm r}(r)-\Sigma_{\rm bg}}{ \Sigma_{\rm bg}}=\frac{\Sigma_{\rm r}(r)}{\Sigma_{\rm bg}}-1\, ,
\end{equation}
where $\Sigma_{\rm r}$ and $\Sigma_{\rm bg}$ are  
the local galaxy, and the background galaxy surface density, respectively. 
The overdensity parameter then equals zero, 
$\delta_{\rm g}=0$ (i.e., $\Delta \Sigma=0$), for no observed overdensity, $\delta_{\rm g}>0$ indicates an overdensity, while $\delta_{\rm g}<0$ indicates an underdensity. 

The average value of the background surface density ($\Sigma_{\rm bg}$) was computed using  nine differently positioned, circular, and not overlapping $A_{\rm bg}=706.86$~arcmin$^2$ areas. Errors, reflecting statistical fluctuations, were then assigned on $\delta_{\rm g}(r)=0$, i.e. $\sigma_{\rm \delta_g=0}(r)$ (see Section\,3.1.3. in \citealt{smolcic17c} for more details). 

The results are shown in the right panel of Figure~\ref{fig:environment_vta}, where we show $\delta_g$ as a function of radius, computed in steps of $0.1$ arcminutes, as well as the signal-to-noise ratio, $\delta_g / \sigma_{\rm \delta_g}$.  We also indicate $\delta_g$ values for which the Poisson probability of observing more or equal  number of sources than expected from the background galaxy surface density is $\leq0.05$, considering these to be robust overdensity values. As seen in Figure~\ref{fig:environment_vta}, a galaxy overdensity is  detected out to a comoving radius of 5\,Megaparsec (Mpc) and it is centered at only $\approx$500 comoving\,kpc away from J1000$+$0234. That extent matches the one expected for massive proto-galaxy clusters \citep[e.g.,][see our Section\,\ref{subsec:fate}]{chiang13}. At  radii below $0.5\,\rm arcmin$/$1.2\,
\rm comoving\,Mpc\;(cMpc)$, in particular, we derive an overdensity parameter  of $6\pm1$. This analysis confirms the results based on the COSMOS2015 photometric redshift catalog and the spectroscopic analysis presented in \citet{smolcic17c}. Moreover, {  this result adds more evidence on the occurrence of LABs in galaxy overdensities at high redshifts}
 \citep[e.g.,][]{matsuda04, alexander16, badescu17, guaita22}. \\

\begin{table*}
	\caption{Ly$\alpha$ line emitters in the vicinity of J1000+0234.}
	
	\label{tab:companions}
	\begin{tabular}{lcccccccc}
		\hline
		\hline
		Companion & RA & DEC &  $\lambda_{\rm central}$ &  $z$ & Proj. area & Iso Flux & Ly$\alpha$ luminosity \\
		& [hh:mm:ss.sss] & [dd:mm:ss.ss]  & [\AA] &   &  [arcsec$^{2}$\,/ \, kpc$^{2}$] & [$\times 10^{-17}$\fcgs] &  [$\times 10^{42}$\lcgs]\\
		\hline
		C1 & 10:00:54.158   & 02:34:22.32  &  6701.4 &   $4.512\pm0.001$ & 7.2 / 315 & $0.74\pm 0.03$ &  $1.51 \pm 0.07$    \\	
		
		C2$^{\dagger}$ & 10:00:55.679   & 02:34:13.23  &  6721.8 &   $4.529\pm0.001$ & 10.5 / 460 & $0.83\pm 0.04$ &  $1.71 \pm 0.10$   \\

		C3 & 10:00:54.398   & 02:34:14.25  &  6732.5 &   $4.538\pm0.002$ & 13.8 / 600 & $0.85\pm 0.04$ &  $1.76 \pm 0.10$   \\
		
		C4 & 10:00:54.105   & 02:34:20.55  &  6733.0 &    $4.538\pm0.002$ & 25.2 / 1090 & $1.46\pm 0.05$ &  $3.03 \pm 1.10$   \\

		C5 & 10:00:54.841   & 02:34:27.49  &  6736.0 &   $4.541\pm0.002$ & 10.7 / 462 & $1.76\pm 0.05$ &  $3.65 \pm 0.10$   \\

		\hline 
	\end{tabular}
	
	$^\dagger$ Assuming that C2 corresponds to the  [O{\sc ii}]\,$\lambda$3727 doublet (see Section\,\ref{subsect:environment} for details), we compute $z=0.803\pm0.001$, proj. area of 580\,kpc$^{2}$, and  [O{\sc ii}] luminosity of $2.36\pm0.11 \times10^{40}$\lcgs. 
\end{table*}

\section{Discussion}\label{sec:discussion}
In this section, we identify the possible ionization sources of the extended \Lya\, emission in J1000+0234. We also explore the \Lya\, and \heii\, line profile across the nebula to infer the  large-scale movements of the gas in the J1000$+$0234 system. Finally, we discuss the evolutionary path that J1000$+$0234 might follow, including the potential quenching of satellite galaxies and the origin of the red sequence of galaxies at $z\approx2$.

\subsection{Ionization source}
\label{subsec:ionization_source}

Four main physical mechanisms have been proposed to explain the origin of circumgalactic \Lya\, emission: (a) recombination radiation following photoionization by   UV photons \citep[usually known as fluorescence;][]{hogan87, cantalupo05}; (b) UV photons from shock-heated gas powered by galactic outflows and/or relativistic jets \citep[e.g.,][]{taniguchi00, arrigoni15}; (c) resonant scattering of \Lya\, photons from an AGN and star formation \citep[e.g.,][]{moller98, villar-martin96,dijkstra09,  hayes11, cantalupo14, mas-ribas17, denbrok20}; and (d) cooling radiation  from gas falling into  dark matter halos \citep[e.g.,][]{fardal01,dijkstra09, daddi21}. The latter is, however, unlikely to be the dominant process in J1000+0234, as there exist (at least) two strong sources of ionization within the nebula. First, there is a low-mass SFG (J1000+0234$-$South) with a UV luminosity that implies a SFR of $148\pm 8$\mpyr. Second, there is a  dust-rich, rotating disk galaxy (J1000+0234$-$North) with a UV-based SFR of $52.6\pm 8.5$\mpyr, which represents $\approx 10\%$ of the total SFR (i.e., IR+UV SFR) of this massive SMG \citep{gomez-guijarro18}. 

Apart from star formation activity,  two radio sources within the Ly$\alpha$ nebula  might indicate AGN activity. The brightest radio source \citep[$S_{\rm 3\,GHz}=25\pm 6 \mu {\rm Jy}$;][]{smolcic17} is centered at the locus of the SMG J1000$+$0234$-$North. We derive a monochromatic 1.4\,GHz radio luminosity\footnote{Derived from the observed  3\,GHz flux density and using the relation $S_{\nu}\propto \nu^{-\alpha}$  with $\alpha=0.7$} of $L_{\rm 1.4\,GHz}=5.2\pm1.2\times 10^{24}\,\rm W\,Hz^{-1}$  that is consistent with that of Type II AGNs at $z\approx 3.1$ \citep{ao17, marques-chaves19}. 
The close-to-orthogonal position angles (PA) of the radio emission ($33\pm6\deg$) and rotating disk of J1000$+$0234$-$North \citep[$145\pm5\deg$;][see our Figure\,\ref{fig:vel-int_lines}]{fraternali21} also hints at radio AGN activity, as the elongated radio emission  suggests the presence of a jet  moving perpendicular to the rotating {  [C{\sc ii}]} disk. {  However, the  FIR-to-1.4\,GHz luminosity ratio $q_{\rm FIR}=1.9\pm0.1$ of J1000$+$0234$-$North is not significantly lower than the $q_{\rm FIR}$ value for SFGs  at $z=4.5$ of $2.1\pm0.2$ \citep{magnelli15, delhaize17},  suggesting that there is not a noticeably excess of radio emission in J1000+0234  due to AGN activity.   }

The faintest radio source within the Ly$\alpha$ nebula \citep[$S_{\rm  3\,GHz}=15\pm2\mu \rm Jy$;][]{smolcic17} is not detected in existing ALMA imaging at 870\,$\rm\mu m$. It is located at $1\farcs5$ to the southeast of J1000$+$0234$-$North, along the direction of the potential radio jet coming out of the massive SMG. In this first scenario, the faint radio source might be tracing a radio jet/lobe. However, because there is faint UV emission at a projected distance  $<0\farcs5$ from the center of the faint radio source, it is also possible this is linked to a faint (and thereby undetected) SFG. Under this assumption, and using the rms noise level of the 870$\rm \,\mu m$ map, we infer a $1\sigma$ upper limit of $q_{\rm FIR}$ of 0.45\footnote{The FIR luminosity is derived by adopting the rms noise level of the ALMA imaging as an upper limit of the 870\,$\rm \mu m$ flux density. We then use the following approximation: $S_{\rm 870\,\mu m}^{\rm SMG}/S_{\rm 870\,\mu m}^{r} \propto L_{\rm FIR}^{\rm SMG}/L_{\rm FIR}^{r}$, where $S_{\rm 870\,\mu m}$ and $L_{\rm FIR}$  are the 870\,$\rm \mu m$ flux density and FIR luminosity of the SMG and the compact radio source ($r$), respectively. Note that in this approximation, we assume that the radio source is linked to a dusty (yet faint and undetected) SFG whose dust spectral energy distribution  has the same properties as that of the bright SMG.}. This  value is  five times lower than the $q_{\rm FIR}$ value for SFGs  at $z=4.5$. As suggested by \cite{radcliffe21}, such a large radio excess with respect to the FIR-radio correlation of SFGs strongly suggests the presence of an AGN. We thus conclude that the faint radio source is likely another source of ionization of the Ly$\alpha$ nebula, either because this is a radio jet/lobe of the massive SMG or because this is tracing an AGN host galaxy that is not robustly detected in the UV nor FIR. Deeper multi-wavelength, high-resolution data is needed to perform robust AGN diagnostics. For instance, current X-ray data only provides an  upper limit for the X-ray luminosity of J1000+0234 of $10^{43.1}\rm erg\,s^{-1}$ \citep{smolcic15}, which is still significantly higher than the standard threshold to select  X-ray AGN \citep[$10^{42}\rm erg\,s^{-1}$;][]{szokoly04}.

\subsubsection{CIV/Ly$\alpha$ and HeII/Ly$\alpha$ line ratios}

After identifying all the potential ionization sources in the field of the extended Ly$\alpha$ blob, we use the \civ\, and \heii\, emission lines  to better identify the origin and physical properties of the emitting gas. 

In Figure\,\ref{fig:line_ratios}, we present the spatial distribution of the C{\sc iv}/Ly$\alpha$ and He{\sc ii}/Ly$\alpha$ surface brightness (SB) ratios (upper panels) and their associated $1\sigma$ error (lower panels). These SB ratios are obtained as follows: 1) we obtain an optimally extracted SB map of \Lya\, emission using a spatial boxcar smoothing of $2\times 2$ pixels, to match the smoothing applied to the C{\sc iv} and He{\sc ii} SB maps (see Figure\,\ref{fig:vel-int_lines}), 2) we regrid the C{\sc iv} and He{\sc ii} SB maps to match both the pixel and sky coordinate system of the \Lya\, SB map, 3) we measure the rms noise level of the C{\sc iv} and He{\sc ii} SB maps, 4) we replace each pixel without detected emission (i.e., $\rm SNR<3$) in the optimally extracted C{\sc iv} and He{\sc ii} SB maps with the $1\sigma$ noise value derived above, 5) we derive the C{\sc iv}/Ly$\alpha$ and He{\sc ii}/Ly$\alpha$ line ratio on a pixel-by-pixel basis. Note that for the regions where \civ\, or \heii\, line emission is not detected, the C{\sc iv}/Ly$\alpha$ and He{\sc ii}/Ly$\alpha$ values correspond to the line ratio $1\sigma$ upper limit. Finally, for the regions where the  C{\sc iv}/Ly$\alpha$ and He{\sc ii}/Ly$\alpha$ line ratios are derived, we estimate the associated $1\sigma$ error using the rms noise levels of the Ly$\alpha$, C{\sc iv}, and He{\sc ii} maps via standard error propagation on a pixel-by-pixel basis. Based on the spatial distribution of \Lya\,, \civ\,, and \heii\, line emission, and their associated ratios (Figure\,\ref{fig:vel-int_lines} and \ref{fig:line_ratios}), we identify three regions of interest within the Ly$\alpha$ nebula: 

\begin{figure*}
	\begin{center}
		\includegraphics[width=11.cm]{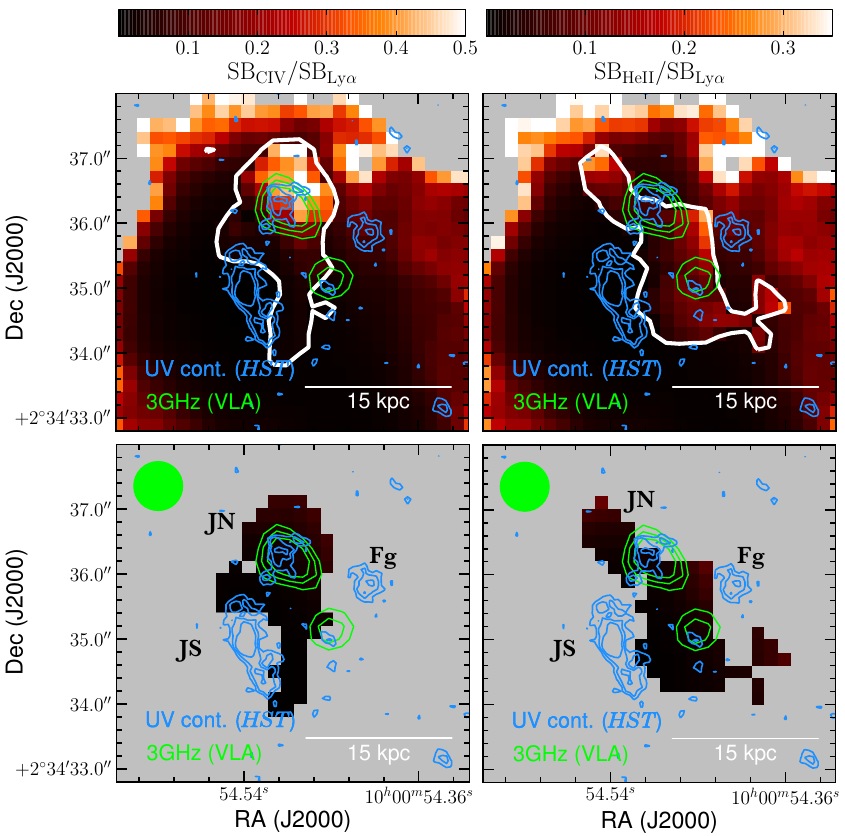}
		\caption{{\it Upper panels:} Spatial distributions of the C{\sc iv}/Ly$\alpha$ and He{\sc ii}/Ly$\alpha$ SB ratios. The white contour delimits the region where C{\sc iv} or He{\sc ii} is detected at a SNR $>3$. {  The blue contours show the rest-frame UV continuum from the \textit{HST}/F160W imaging and the lime contours display the VLA 3\,GHz radio continuum emission \citep{smolcic17}. Contour levels are the same as in Figure\,\ref{fig:vel-int_lines}. } The pixel values outside the white contours are 1$\sigma$ upper limits to the C{\sc iv}/Ly$\alpha$ and He{\sc ii}/Ly$\alpha$ SB ratios (see text for details).  {\it Lower panels:} Spatial distribution of the 1$\sigma$ uncertainties of the C{\sc iv}/Ly$\alpha$ and He{\sc ii}/Ly$\alpha$ SB ratios. {  The VLA beam shape is shown in the upper-left corners.} The regions with the highest C{\sc iv}/Ly$\alpha$ and He{\sc ii}/Ly$\alpha$ ratios are located in the vicinity of J1000$+$0234$-$North (JN): a massive SMG, with signatures of AGN activity, that has a UV and 3\,GHz radio counterpart. The lowest C{\sc iv}/Ly$\alpha$ and He{\sc ii}/Ly$\alpha$ SB ratios, with a 1$\sigma$ upper limit of $\approx0.003$,  are located at the locus of J1000$+$0234$-$South (JS): a low-mass SFG that is bright in the UV and no signatures of AGN activity.  } 
		\label{fig:line_ratios}
	\end{center}
\end{figure*}

\begin{figure}
	\begin{center}
		\includegraphics[width=7cm]{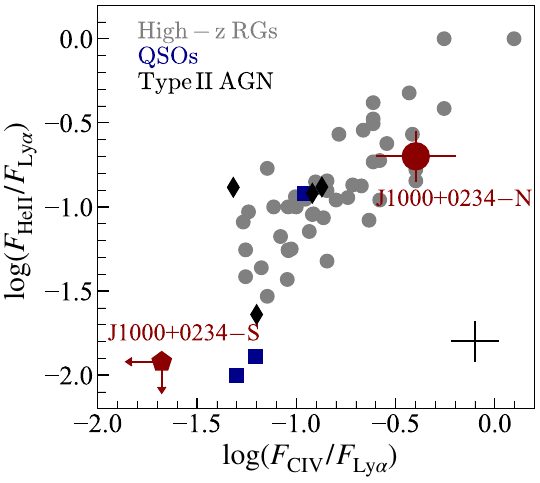}
		\caption{Ly$\alpha$, He{\sc ii}, and C{\sc iv} Line ratios of LABs previously associated with high-redshift radio galaxies (High-z RGs), QSOs, and Type\,II AGN \citep[using data from][and references therein]{marques-chaves19, marino19, wang21}. The typical length of the error bars is shown in the bottom-right corner.    The average line ratios around J1000$+$0234$-$North, the massive SMG with a potential AGN, are  consistent with those observed in  High-z RGs and Type\,II AGN. On the contrary, the 3$\sigma$ upper limits  to the average line ratios across the extent of J1000$+$0234$-$South, the low-mass starburst, suggest that a different mechanism is driving \Lya\, emission here.  }
		\label{fig:comparison_lineratios}
	\end{center}
\end{figure}

\begin{enumerate}

    \item  First, we identify the area surrounding the  massive, dusty, disk galaxy J1000+0234$-$North where \civ\,  emission is maximal. This indicates the  presence of a metal-enriched gaseous halo \citep[e.g.,][]{dopita07, reuland07, kolwa19} around this source,  which is consistent with J1000+0234$-$North being a massive, dust-rich, and --likely-- a metal-enriched system \citep[given the mass-metallicity relation; see][for a review]{maiolino19}. The detection of \civ\, also implies the presence of  high-ionization radiation to triply ionize carbon, making this line a potential tracer of AGN activity \citep{baskin05, mignoli19}. The average CIV/Ly$\alpha$ line ratio around the vicinity of the SMG is $0.40\pm0.15$.  This is significantly higher than { most of the} line rations found in non-AGN LBG spectra \citep[CIV/Ly$\alpha\leq 0.02$;][]{shapley03}, but more consistent with those values found in the LBG harboring a narrow-line AGN \citep[$\rm CIV/Ly\alpha\approx0.25$;][]{shapley03} and  local Seyfert galaxies \citep[$\rm CIV/Ly\alpha\approx0.21$;][]{ferland86}. This finding thus strengthens the conclusion of our radio-FIR analysis that J1000+0234$-$North harbors an AGN.  
    
    \item Second, we recognize a more extended region where \heii\, emission is detected (Figure\,\ref{fig:line_ratios}). This is elongated along the relative position of the two radio continuum sources. The regions where the He{\sc ii}/Ly$\alpha$ ratio is maximal  are placed on either side of J1000+0234$-$North, roughly along the direction of the bright radio source's major axis. Such a high line ratio in these regions, averaging He{\sc ii}/Ly$\alpha=0.20\pm0.07$,  is close to the recombination case for fully ionized hydrogen and helium
    \citep[HeII/Ly$\alpha=0.3$ for Case B  recombination][]{cantalupo19}. This suggests that,  in the two nearby regions on either side of J1000+0234$-$North, the observed He{\sc ii}/Ly$\alpha$ line ratio is consistent with recombination processes, which take place after the exposure of the circum-galactic medium to the UV-flux from the AGN in J1000+0234$-$North. With a 1.4\,GHz radio luminosity of $\rm 5.2\pm1.2\times 10^{24}\,W\,Hz^{-1}$, J1000+0234$-$North might harbor an AGN bolometric luminosity of the order of $10^{39.5}\,\rm W\,Hz^{-1}$ \citep[see Figure\,3 of][]{sulentic10} that is enough to produce an  ionizing photon flux of $\phi \sim 10^{55}\,\rm photons\,s^{-1}$ \citep[see Section\,3.3.1 of][]{valentino16}. In the case of photoionization \citep[see Section\,3.1 of][]{cantalupo17}, such a photon flux implies a \Lya\, line luminosity of the order of $10^{44}\,\rm erg\,s^{-1}$ that is comparable with the measured value. Additionally, the unobscured UV emission from J1000+0234$-$North might also account for a fraction of the total ionizing photons budget.  Considering  the relation ${\rm SFR}/M_\odot\,{\rm yr^{-1}}=0.62 \times L_{\rm Ly\alpha}/10^{42}\,\rm erg\,s^{-1}$ \citep{ao17}, 
        the unobscured,  UV-based SFR of J1000$+$0234$-$North {  \citep[reported by][]{gomez-guijarro18}} leads to a \Lya\, line luminosity of $\rm 33\pm5 \times 10^{42}\,erg\,s^{-1}$ that is a factor 0.8 the measured value.  Thus, an AGN  and unobscured star formation activity in J1000+0234$-$North  should produce enough ionizing photons to  drive the Ly$\alpha$ nebula around the J1000+0234 system.
    Moreover, the tentative evidence of a radio lobe/jet, which matches the locus of bright \heii\, and faint UV continuum emission (Figure\,\ref{fig:vel-int_lines}),  hints at the possibility of jet-cloud interactions \citep[e.g.,][]{maxfield02, humphrey06, steinbring14} that could represent another ionization source in this region. In fact, the average C{\sc iv}/Ly$\alpha$ and HeII/Ly$\alpha$ line ratios in the vicinity of J1000+0234$-$North of 0.4 and 0.2, respectively, are consistent with those observed in Type II AGNs and high-redshift radio galaxies (Figure\,\ref{fig:comparison_lineratios}).

    \item Third, there is a region where \Lya\, emission is concentrically distributed around the low-mass SFG J1000$+$0234$-$South. This is also the area where \Lya\, emission peaks, which strongly suggests that  \Lya\, emission is driven by photoionization \citep{villar-martin07b} from the young starburst in J1000-0234$-$South, complementing the incident ionizing radiation from the nearby AGN in J1000+0234$-$North. 
        The   UV-based SFR of J1000$+$0234$-$South {  \citep[reported by][]{gomez-guijarro18}} implies \Lya\, line luminosity of $\rm 92\pm5 \times 10^{42}\,erg\,s^{-1}$ that is two times the measured value, which suggests that the observed \Lya\, luminosity could also be explained by the intense star formation in J1000-0234$-$South.     No \civ\, nor \heii\, line emission is detected across the UV extent  of J1000+0234$-$South, which only allows us to estimate a $1\sigma$ upper limit of the  C{\sc iv}/Ly$\alpha$ and He{\sc ii}/Ly$\alpha$ line ratio of $0.007$ and $0.004$, respectively. 
        Assuming that the starburst in J1000+0234$-$South  does not produce a sufficiently hard radiation field, such low ratios suggest that the \Lya\, emitting gas around this low-mass galaxy does not receive enough incident flux above 4 Rydberg from the AGN in J1000$+$0234$-$North to double-ionize helium, either because of obscuration in that particular direction or because this part of the nebula is at a large distance from the AGN in J1000$+$0234$-$North. Indeed, the redshift from the \heii\, line of $4.545\pm0.002$, tracing the gas in the vicinity of the AGN,  is mildly higher than the photometric redshift of  J1000+0234$-$South ($4.476\pm0.025$).
However, as we discuss in the following section,  due to   the  --likely high-- peculiar velocities in this complex system, we can not disentangle the physical separations (along the line of sight) of all the sources within the extended Ly$\alpha$ blob of J1000+0234.
        Alternatively,
        these rather low  He{\sc ii}/Ly$\alpha$ ratio across and around J1000+0234$-$South can  be a result of
a broad density distribution of the circum-galactic gas in this region. 
Photoionization models for case B recombination from \citet[][]{cantalupo19}, that adopt a lognormal density distribution, suggest that  large lognormal dispersions ($\sigma_{\rm gas}$) of the gas density produce both smaller He{\sc ii}/Ly$\alpha$ line ratios and large \Lya\, SB values. Here, we observe  $\rm SB_{\rm Ly\alpha}\gtrsim 10^{17} \rm erg\,s^{-1}\,cm^{-2}\,arcsec^{-2}$ and  He{\sc ii}/$\rm Ly\alpha<0.004$ across and within the vicinity of the UV extent of J1000+0234$-$South, which imply  $\sigma_{\rm gas}$ values above 2.0  \citep[][see their Figure\,8]{cantalupo19} that are comparable with those observed in the interstellar medium (ISM) of galaxies.

\end{enumerate}

In summary, based on their spatial distribution and potential ionization sources, we have identified three relevant zones within the Ly$\alpha$ nebula of J1000+0234. 1) A region around J1000$+$0234$-$North where  C{\sc iv} is present and He{\sc ii}/Ly$\alpha$ ratio is relatively high, this is likely where a Type II AGN is. 2) A region in the vicinity of the faint radio source  with a relatively high He{\sc ii}/Ly$\alpha$ ratio that is possibly driven by jet-cloud interactions. 3) A region corresponding to the \Lya\, peak and J1000+0234$-$South. This is likely too distant and too dense for the AGN to double ionize He. Here, \Lya\, emission might be also due to the local (stellar) radiation field.

\begin{figure*}
\begin{center}
		\includegraphics[width=13.5cm]{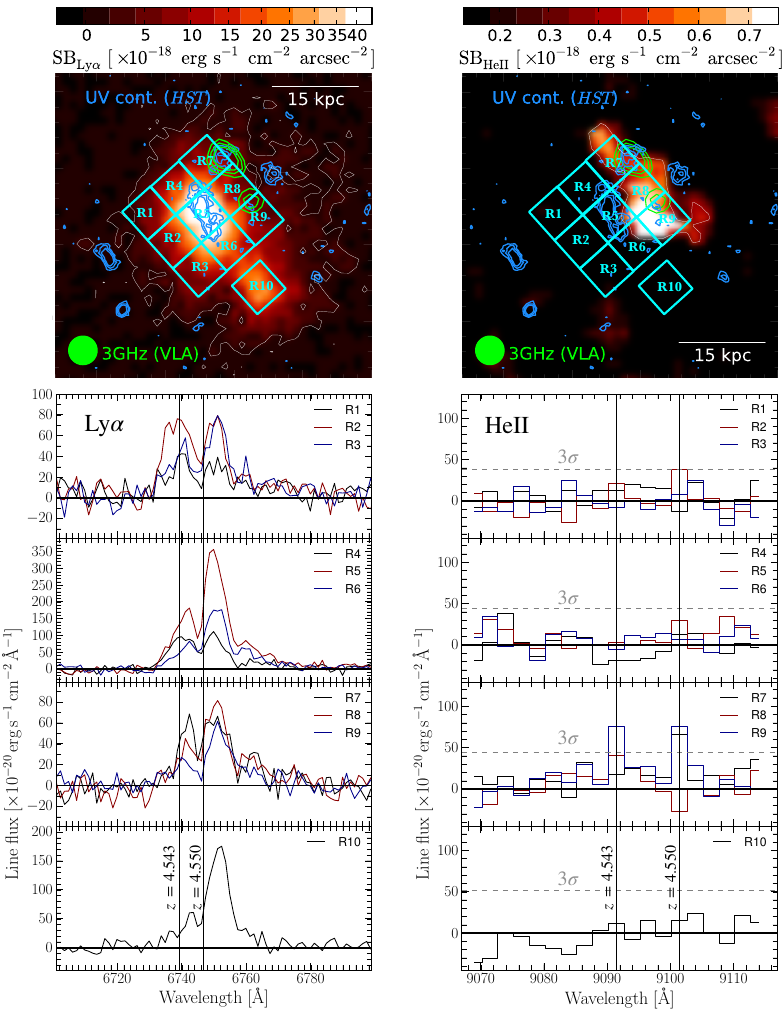}
		\caption{ {\it Upper panels:} Optimally-extracted SB maps of \Lya \, ({\it left}) and \heii\,({\it right}) emission in J1000$+$0234.  The thin white contours indicate the SNR = 3 isophote. The blue contours show the rest-frame UV continuum from the \textit{HST}/F160W imaging and the lime contours display the VLA 3\,GHz radio continuum emission \citep{smolcic17}. {   Contour levels are the same as in Figure\,\ref{fig:vel-int_lines}. The VLA beam shape is shown at the bottom-left corners.} The cyan squares delimit the regions from which the 1D spectra shown in the lower panels are obtained. {\it Lower panels:} \Lya \, ({\it left}) and \heii\,({\it right}) line profiles across the extent of the J1000$+$0234 nebula. The \heii\, spectra are averaged over a 3\,\AA-width bin. The horizontal, grey, and dashed lines illustrate the typical $3\times$ noise rms level of the 1D \heii\, spectra.  The black vertical lines show the expected central wavelength of the \Lya\, and \heii\, lines of a source at $z=4.543$ and $z=4.550$, respectively. The double-peaked nature of the \Lya\, line persists across the brightest \Lya-emitting regions. The central wavelengths of the two peaks do not significantly vary across the 10 regions explored here. 
		\heii\, line emission is only detected in regions R7 and R9, corresponding to the locus of the bright SMG and the site of a possible radio jet (see details in Section\,\ref{subsec:ionization_source}). The marginal evidence of  two peaks in the (non-resonant) \heii\, line profile of regions R7-R9 suggests the presence of two different clouds along the line of sight.     }
		\label{fig:kinematics_1}
\end{center}
\end{figure*}

\subsection{Ly$\alpha$ kinematics}
\label{subsec:kinematics}
The 1D spectrum of the J1000+0234 system (Figure\,\ref{fig:lines}) reveals a wide, multi-peaked, and asymmetric \Lya\, line profile that hints at complex and perturbed gas kinematics of a nebula or multiple structures with distinct peculiar velocities. To disentangle the contributions from these  scenarios we combine information from our \Lya\, emission map with our \heii\, line detection.

At first glance, the integrated \Lya\, line profile  of the J1000$+$0234 nebula {  (Figure\,\ref{fig:lines})} resembles the expected profile of an expanding thin shell of neutral gas:   one small blue-shifted wing, one narrow emission peak centered nearly at the systemic velocity, a red-shifted peak, and one  extended red wing \citep[e.g., ][]{verhamme06, orsi12}. According to this model, the wide and asymmetric \Lya\, line profiles are a result of galactic-scale  outflows driven by a central source  \citep[e.g., ][]{taniguchi00, dawson02, dijkstra09}. While photons experiencing multiple backscatterings produce an extended and redshifted wing,   the external and expanding neutral H{\sc i} gas  along the line of sight absorbs the blue velocity component. This highly idealized model of a  galaxy-scale expanding shell of neutral hydrogen around a central source {seems}, however,  
{  insufficient to explain the multi-component and asymmetric morphology  of the nebula and the multiple ionizing sources in the system (including AGN and star formation, see Section\,\ref{subsec:ionization_source}).} 

{ 
{Another} possible explanation for the asymmetric double-peaked Ly$\alpha$ line profile of J1000$+$0234 is the absorption of blue-shifted emission by the intervening intergalactic medium \citep[e.g.,][]{laursen11}. \citet{hayes21}, in particular, find that the  ratio of the blue/red wing intensity  of a double-peaked Ly$\alpha$ line at $z=4.5$ is only $\approx0.1$ (see their Figure\,9). By extracting and inspecting the \Lya\, line profile at different spatial locations of the J1000$+$0234 nebula (see left panels of Figure\,\ref{fig:kinematics_1}), we find that the ratio of the blue/red wing intensity is larger than 0.1 and can be up to 1. Under this scenario, the  low foreground absorption of the blue peak could indicate higher ionizing radiation from the J1000$+$0234 components that diminishes the neutral H{\sc i} gas budget.  } 

\begin{figure*}
	\begin{center}
		\includegraphics[width=17.6cm]{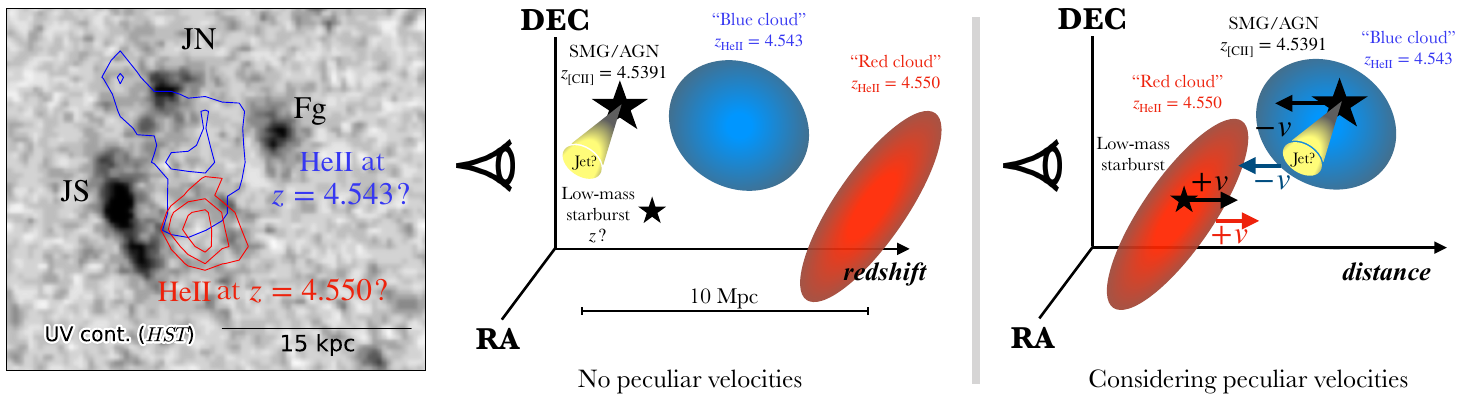}
		\caption{({\it Left:}) Spatial distribution of redshifted/blueshifted  \heii \, emission of  J1000$+$0234 (blue and red contours, respectively) overlaid on the {\it HST}/F160W continuum image, which shows the rest-frame UV emission from the SMG J1000$+$0234$-$North (JN) and the low-mass starburst J1000$+$0234$-$South (JS). The prominent continuum source to the right of the J1000$+$0234 system is a foreground (Fg) galaxy \citep{capak08}. The red and blueshifted emission maps are obtained using the same 3D masks obtained during our source extraction procedure, but considering the channels lower/higher than 9095.0\,\AA\, to get optimally-extracted NB images of the blue/redshifted \heii\, emission. The contour levels are at 3, 4, 5, 6, and 7 times the noise rms level.  The spatial (and spectral) offset (see Section\,\ref{subsec:kinematics}) between the blueshifted and redshifted  \heii \, emission of J1000$+$0234 suggests that these might represent two different clouds in the J1000$+$0234 system at $z=4.543\pm0.002$ and $z=4.550\pm0.002$. ({\it Middle:})  Cartoon illustrating the configuration along the line of sight of multiple components in the J1000$+$0234 system as inferred from their redshifts.  If the effect of peculiar velocities is neglected,  the redshift values would imply unrealistic comoving radial distances between the components on the order of Mpc (see Section 5.2 for details). ({\it Right:}) Assuming that the redshift values are also driven by peculiar velocities in the J1000$+$0234 system, that  is in the center of a galaxy overdensity, we propose that the  two emitting clouds are on a  collision course {  or receding from each other (after a passage)} with a relative velocity of $\approx400$\,\kmps. {  In this cartoon, we illustrate the former scenario where the components are  approaching one another before the passage}. The lower [C{\sc ii}]-based redshift of J1000$+$0234$-$North could be a result of a larger peculiar negative velocity along the line of sight. The direction and magnitude of the  peculiar velocities of all the relevant components are represented by the black, blue, and red arrows.  
 }
		\label{fig:kinematics_2}
	\end{center}
\end{figure*}

Alternatively, the two prominent peaks of the \Lya\; line profiles could correspond to two emitting clouds along the line of sight. We evaluate this scenario by exploring the non-resonant \heii\,emission extracting  1D spectra from the same regions employed to analyze the \Lya\, line profile across the nebula (see the right panels of Figure\,\ref{fig:kinematics_1}). \heii\, emission is only detected in regions R7 and R9, i.e., around the SMG and the potential radio jet (see Section\,\ref{subsec:ionization_source}). Interestingly, the separation between the potential peaks of the He\,{\sc ii} line profile in R7/R9 is $10$\,\AA, which  corresponds to a velocity shift of $\approx 330$\,\kmps. These values are similar to the velocity offset we observe in the double-peaked \Lya\, line profile across the entire nebula of $\approx 400$\,\kmps. Hence,  the peaks at 9091.5 and 9101.5 \AA\, in the He\,{\sc ii} line profile of R7 and R9 might represent two different emitting clouds along the line of sight with redshifts $4.543\pm 0.002$ and $4.550\pm0.002$. We note that the expected central wavelength of the \Lya\, line profile from those potentially distinct regions, shown in the lower-left panel of Figure\,\ref{fig:kinematics_1}, does not agree with the observed central wavelength of the peaks, which could be explained by the resonant nature of the \Lya\, line emission that tends to be redshifted against the radial systemic velocity. Moreover, the scenario of two distinct clouds contributing to the \Lya\, nebula is reinforced by the significant spatial offset between the red and blueshifted \heii\, emission of $\approx 6$\,kpc  (see Figure\,\ref{fig:kinematics_2}).

Under the scenario of two different line emitting clouds in the J1000$+$0234 system, the observed wavelength difference between the line peaks in the \Lya\, and \heii\, line profiles could be a result of peculiar velocities and/or a different distance between them. If the velocity difference ($\delta_{v}$) is solely due to the Hubble flow, the peaks shift would imply a physical distance of $\delta_{v}/H_{0}\sim 5$\,Mpc. This value is comparable to the extent of the galaxy overdensity within which the J1000$+$0234 system resides (see Section\,\ref{subsect:environment}). We, therefore, consider this scenario unlikely. Instead, because J1000$+$0234 is a pair of likely interacting galaxies near the center of a galaxy overdensity, where peculiar velocities are on the order of a few hundred  \kmps \citep[e.g.,][]{ceccarelli05, santucho20}, we would expect that the observed velocity offsets are largely driven by peculiar velocities. In this context, and based on the blue and red peak of the  \heii\, and \Lya\, lines, the apparent two emitting clouds in the J1000+0234 system  are approaching one another {  (or receding from each other after a passage)} with a relative velocity of $\approx400$\,\kmps. 
The relative motion between the multiple components in the J1000$+$0234 system could also explain the disagreement between the [C{\sc ii}]-derived redshift of J1000$+$0234$-$North \citep[$4.5391 \pm 0.0004$;][]{fraternali21} and the redshift of the \heii-emitting clouds ($z\approx 4.545$). 
A  plausible scenario is that J1000$+$0234$-$North has a larger peculiar (negative) velocity --along the line of sight-- than the blue-shifted cloud within which the SMG/AGN is embedded (see right panel of Figure\,\ref{fig:kinematics_2}). Since there are no spectroscopic redshift measurements of J1000$+$0234$-$South, we can only hypothesize that this low-mass starburst is embedded in the red-shifted cloud and is {  approaching  (or receding from)} the blue-shifted cloud and the massive SMG J1000$+$0234$-$North.

\subsection{The nature and fate of the complex J1000+0234 system}
\label{subsec:fate} 
The J1000+0234 complex (J1000+0234$-$North and J1000+0234$-$South) and its companions   reside in the central region of a large-scale overdensity of galaxies at $z=4.5$, supporting the role of LABs as tracers of rich environments at high redshift \citep[e.g.,][]{matsuda04, alexander16, badescu17, herez20, guaita22}.  In these overdense regions, the evolution of galaxies might be highly affected by environmental processes. Certainly,  \citet{gomez-guijarro18}  proposed that J1000+0234$-$North and J1000-0234$-$South are undergoing a minor merger with a stellar mass ratio of 1:10, which might have contributed to triggering a Type II AGN and a vigorous starburst in J1000+0234$-$North -- as suggested by numerical simulations \citep[e.g.,][]{hopkins06}. Merger activity could also explain the apparent elongation of both the rest-frame UV and \Lya\, emission around J1000+0234$-$South (Figure\,\ref{fig:vel-int_lines}),  hinting at a tidal structure of young stars and  gas resulting from the gravitational interaction. While such a merger could be stripping gas from the minor companion J1000+0234$-$South, it is known that the mechanisms leading to stripping can briefly enhance  star formation \citep[e.g.,][and references therein]{cortese21}, consistent with the starburst nature of J1000+0234$-$South. As discussed in Section \ref{subsec:ionization_source},  vigorous star formation activity might be the dominant ionization source of \Lya\, emission around  J1000+0234$-$South, while AGN activity and  dust-unobscured star formation are  the potential main drivers of extended \Lya\, emission around J1000+0234$-$North.  Our kinematical analysis suggests the presence of two HeII emitting clouds along the line of sight  with a relative velocity of 400\,\kmps, consistent with the scenario in which J1000+0234 is {  an interacting system whose components are on a collision course or receding from each other (after a passage)}. Interestingly, despite the indications of galaxy interactions in this system, the SMG J1000+0234$-$North preserves a rotationally-supported gas disk, suggesting that any previous merging activity (that triggered the AGN and the starburst phase in the J1000$+$0234 system)  involved low-mass companions that did not disrupt the disk. This scenario has been proposed for the  SMG W0410-0913 that also harbors a rotationally-supported gas disk and resides in an overdense environment at $z\approx 3.6$ \citep{ginolfi22}. The results presented here [and in \citet{ginolfi22}] agree with expectations from numerical simulations indicating that, rather than major mergers, smooth accretion from cosmological filaments and minor mergers are the dominant mechanisms regulating the growth of high-redshift galaxies in overdense environments \citep{romano-diaz14}. Also, {  the observed properties of J1000+0234 seem to} agree with numerical simulations predicting that merging activity does not necessarily {  disrupt disks}, particularly in gas-rich systems like J1000+0234 \citep{schinnerer08} where even the residual gas from major mergers may reform a disk \citep[e.g.,][]{springel05, romano-diaz11, martin18, peschken20} ---albeit other physical conditions like  the orbital configuration of the merging system  \citep{zeng21} and {  the time after the merger} may be at play here as well.

The evolutionary path that the system J1000+0234 will follow can be discussed in terms of cluster formation and the assembly of local elliptical galaxies. We first use the predictions from cosmological simulations in \citet{chiang13} to verify if J1000+0234 matches the expected observational signatures of galaxy proto-clusters.
Based on the results shown in Figure\,\ref{fig:environment_vta}, the overdensity around J1000+0234 extends out to a 5cMpc radius (10cMpc diameter) where {  $\delta_{\rm g}\approx0.5$}. Comparing with the results from \citet{chiang13} for a (15\,cMpc)$^{3}$ volume, that is the one that resembles the most to our observations, we find that the overdensity around J1000+0234 at $z=4.5$ has a {  low  probability} ($<20$\%) to evolve into a galaxy cluster with a total mass of $\approx 10^{14}\,M_\odot$ at $z=0$. {  In that event, the J1000$+$0234 system}  would be  consistent with the {  proposed} evolutionary link between $z>3$ SMGs and their descendant, massive elliptical galaxies at $z=0$ \citep{geach05, toft14, stach21}  that preferentially reside at the center of galaxy clusters. Such a scenario
involves 
the abrupt  cessation of star formation  in $z>3$ dusty starbursts \citep[][]{jones17}  that subsequently merge with \emph{quenched} satellite galaxies \citep[e.g., ][]{dave17}. 
{  As previously discussed by \citet{gomez-guijarro18} and \citet{fraternali21}, there exists indirect evidence suggesting a possible evolution of J1000$+$0234 into a local early-type galaxy.}
First, the expected growth in mass and size of the SMG J1000$+$0234$-$North, including the contribution of minor mergers, broadly agrees with the properties of the compact quiescent galaxies at $z\approx2$ \citep{gomez-guijarro18} that are the progenitors of local massive ellipticals.
Second, the inner potential well of the  SMG J1000$+$0234$-$North, probed by the [C{\sc ii}] kinematics, is comparable with that of the most massive early-type galaxies in the local universe \citep{fraternali21}, providing dynamical evidence of the evolutionary link between the J1000$+$0234 system and local ellipticals. 

While the observational results presented here suggest that J1000+0234 reside at the center of a potential proto-cluster,  the mechanisms that can suppress star formation in the  central  SMG J1000+0234$-$North remain unclear. A plausible scenario might involve "mass quenching". Galaxies with stellar masses a factor $\sim 2$ larger than J1000+0234$-$North (i.e., $M_\star\sim 10^{10.5}\,M_\odot$, with dark matter halos  $\sim 10^{12}\, M_\odot$) are expected to form a hot gas corona that can shut off the gas supply from the cosmic web and prevent the cooling of gas \citep[][and references therein]{gabor10}. The activity of an AGN, like the one inferred for J1000+0234$-$North, can  act as an additional heating source \citep[e.g.,][]{croton06}, suppressing the formation of stars in such massive galaxies.

\section{Conclusions}
\label{sec:conclusions}
We use MUSE observations to characterize extended \Lya,
 \civ\, and \heii\, emission around J1000+0234; a pair of  galaxies at $z=4.5$ in the COSMOS field. J1000+0234$-$North  is a massive SMG ($\log(M_\star/M_\odot)=10.1\pm0.1$) exhibiting a rotating [C{\sc ii}] disk  with  $\rm SFR=500^ {+1200}_{-320}$\,\mpyr \citep{gomez-guijarro18, fraternali21}, while  J1000+0234$-$South is a UV-bright, low-mass SFG ($\log(M_\star/M_\odot)=9.2\pm0.1$) with $\rm SFR=148\pm8$\,\mpyr \citep{gomez-guijarro18}. By combining MUSE observations with existing ALMA and {\it HST} data we find the following.

\begin{itemize}
    \item We detect \Lya\, emission that peaks at the locus of J1000+0234$-$South and extends over a projected area of 1853$\,\rm kpc^{2}$, with a  maximum linear projected size of 
    $\approx$43\,kpc (Figure\,\ref{fig:vel-int_lines}). Based  (optimally-extracted) flux of $F_{\rm Ly \alpha}^{\rm obs}=(19.79\pm 0.15) \times 10^{-17}$\fcgs\, we compute a total line luminosity of  $L_{\rm Ly \alpha}^{\rm obs}=(40.17\pm 0.31) \times 10^{42}$\lcgs. J1000+0234 is among the most compact and least luminous high-redshift Ly$\alpha$ nebulae (see Figure\,\ref{fig:comparison_size-vs-llya}).\\

    \item  \heii\, line emission is also detected around J1000+0234$-$North (Figure\,\ref{fig:vel-int_lines}). This emission, extending out to a projected area of $\rm 3.7\,arcsec^2\approx159\,\rm kpc^2$, reaches a maximum {  near} the locus of a faint radio source at $1\farcs5$ to the South-West of J1000+0234$-$North (likely a radio jet). With a flux of   $F_{\rm HeII} ^{\rm obs}=(0.27\pm0.03)\times 10^{-17}$\,\fcgs\, we compute a to total line luminosity of $L_{\rm HeII}=(0.57\pm0.06)\times 10^{42}$\,\lcgs.\\ 
    
    \item We detect \civ\,  line emission around J1000+0234$-$North that extends out to a projected area of $\rm 3.7\,arcsec^2\approx159\,\rm kpc^2$ (Figure\,\ref{fig:vel-int_lines}). The measured flux of $F_{\rm CIV}^{\rm obs}=(0.37\pm0.03)\times 10^{-17}$\,\fcgs\,  leads to a total line luminosities of $L_{\rm CIV}=(0.76\pm0.07)\times 10^{42}$\,\lcgs. \\

    \item We detect three \Lya\, emitters (C3, C4, C5) spanning over a redshift bin $\Delta z \leq 0.007$ (i.e., $\lesssim 380\,\rm km\,s^{-1}$) located at  $\lesssim 140\,\rm kpc$ from J1000+0234 (Figure\,\ref{fig:environment}). These companions have a projected area ranging from 460 to 1090\,$\rm kpc^{2}$ and \Lya\, line luminosities of $1.8-3.6\times 10^{42}$ \lcgs. We identify a galaxy overdensity   that is centered at only $\approx 500$ comoving\,kpc away from the J1000$+$0234 system.  We estimate an overdensity parameter $(\delta_g)$ of {  $6\pm1$} at a comoving radius of {  1.2\,Mpc}. This result supports earlier indications that J1000+0234 lies near the center of a Mpc-scale galaxy overdensity at $z=4.5$ \citep{smolcic17c}. \\
    
    \item The C{\sc iv}/Ly$\alpha$ and He{\sc ii}/Ly$\alpha$ Line ratios are maximal in the vicinity of J1000+0234$-$North (Figure\,\ref{fig:line_ratios}). These line ratios ($\approx 0.2-0.4$) are consistent with  those observed in Type II AGN and high-redshift radio galaxies \citep{shapley03, marques-chaves19}. Here, \Lya\, emission might be driven by the AGN and associated jet in J1000+0234$-$North.\\
    
    \item At the locus of J1000+0234$-$South, where \Lya\, emission peaks but no \heii\, nor \civ\, emission is detected (Figure\,\ref{fig:vel-int_lines} and \ref{fig:line_ratios}), the C{\sc iv}/Ly$\alpha$ and He{\sc ii}/Ly$\alpha$ line ratios are smaller than 0.007 and 0.004, respectively. These low line ratios hint at the possibility that the gas around J1000+0234$-$South does not receive enough incident flux from the AGN. Here, \Lya\, emission might be driven by the starburst in J1000+0234$-$South.   \\

     \item There is marginal evidence for two He{\sc ii}-emitting clouds separated by 10 \AA\, and 6\,kpc (projected on the sky plane). This suggests that the velocity structure of this LAB is the result of two overlapping halos. The two \Lya\, peaks might correspond to two emitting clouds in the J1000$+$0234 system approaching one another {  (or receding from each other after a passage)} with a relative velocity of $\approx 400$\,\kmps. 
     
\end{itemize}

This study strengthens the role of LABs as  observational signatures of galaxy over-densities at high redshifts, within which galaxy mergers can trigger intense star formation and AGN episodes in a central, dust-enshrouded massive galaxy that  preserves a rotationally supported gas disk.  The clustering around J1000$+$0234 also {hints at}  the proposed evolutionary link between SMGs in rich environments and local elliptical galaxies that reside at the center of galaxy clusters.    Moreover, this work is an instructive example of how   spatially-resolved observations of \heii\, and \civ\, line emission  are key to unveiling the driving mechanism of high-redshift LABs, particularly those associated with the core of galaxy overdensities where total line ratios can be strongly affected by blending. This study also showcases the benefits of using the \heii\, line emission to study the gas kinematics within LABs, because observations of the non-resonant \Lya\, line convey more uncertainties to the already complex configuration  of peculiar velocities in the center of galaxy overdensities.  In this context,  observations with the {\it James Webb Space Telescope} will be necessary to map the kinematical structure of $z\approx 4$ proto-cluster of galaxies using, for example, H$\alpha$ and H$\beta$, which will also serve to unveil the relative contribution of AGN, star formation, and cooling radiation to the origin of LABs.

\section*{Acknowledgements}
We thank the reviewer for their constructive comments and suggestions. 
 EFJA gratefully acknowledges the support from the  NRAO staff that made remote working feasible during the COVID-19 pandemic. E.F.J.A,  B.M., and E.R.D.   acknowledge  support of the Collaborative Research Center 956, subproject A1 and C4, funded by the Deutsche Forschungsgemeinschaft (DFG). SC gratefully acknowledges support from the European Research Council (ERC) under the European Union’s Horizon 2020 research and innovation programme grant agreement No 864361. The Cosmic Dawn Center (DAWN) is funded by the Danish National Research Foundation under grant No. 140. Based on observations collected at the European Southern Observatory under ESO programs 0102.A-0448 and 0103.A-0272.

\section*{Data availability}
The data underlying this article can be accessed from the ESO Archive Science portal (\url{http://archive.eso.org/scienceportal/home}) under the programs 0102.A-0448 and 0103.A-0272 with the object identifier ``COSMOSz4'' {  and ``J1000+0234''}. The derived data generated in this research will be shared on reasonable request to the corresponding author.



\bibliographystyle{mnras}
\bibliography{jimenezandrade+22} 




\appendix



\bsp	
\label{lastpage}
\end{document}